\documentclass[twocolumn,useAMS,usenatbib]{mn2e}
\usepackage{graphicx}
\usepackage{natbib}

\usepackage{bm}
\usepackage{amsfonts}
\usepackage{latexsym}
\usepackage[latin1]{inputenc}
\usepackage{amsmath}
\usepackage{epsfig}
\usepackage{amsbsy}
\usepackage{psfrag}
\usepackage{mnfixes}

\newcommand{\f}{\mathrm f}

\newcommand{\n}{\mathrm n}
\newcommand{\p}{\mathrm p}

\newcommand{\x}{\mathrm x}

\newcommand{\crt}{\mathrm c}

\newcommand{\veps}{\varepsilon}

\def \nn  {\nonumber}

\def \eps{\epsilon}
\def \veps{\varepsilon}

\newcommand{\beq}{\begin{equation}}
\newcommand{\eeq}{\end{equation}}

\newcommand{\Bav}{\langle B \rangle }

\def\jnl@style{\rm}
\def\aaref@jnl#1{{\jnl@style#1}}

\def\aaref@jnl#1{{\jnl@style#1}}

\def\aj{\aaref@jnl{AJ}}                   
\def\apj{\aaref@jnl{ApJ}}                 
\def\apjl{\aaref@jnl{ApJ}}                
\def\apjs{\aaref@jnl{ApJS}}               
\def\apss{\aaref@jnl{Ap\&SS}}             
\def\aap{\aaref@jnl{A\&A}}                
\def\aapr{\aaref@jnl{A\&A~Rev.}}          
\def\aaps{\aaref@jnl{A\&AS}}              
\def\mnras{\aaref@jnl{MNRAS}}             
\def\prd{\aaref@jnl{Phys.~Rev.~D}}        
\def\prc{\aaref@jnl{Phys.~Rev.~C}}        
\def\prl{\aaref@jnl{Phys.~Rev.~Lett.}}    
\def\qjras{\aaref@jnl{QJRAS}}             
\def\skytel{\aaref@jnl{S\&T}}             
\def\ssr{\aaref@jnl{Space~Sci.~Rev.}}     
\def\zap{\aaref@jnl{ZAp}}                 
\def\nat{\aaref@jnl{Nature}}              
\def\aplett{\aaref@jnl{Astrophys.~Lett.}} 
\def\apspr{\aaref@jnl{Astrophys.~Space~Phys.~Res.}} 
\def\physrep{\aaref@jnl{Phys.~Rep.}}      
\def\physscr{\aaref@jnl{Phys.~Scr}}       

\title[ QPOs in superfluid magnetars]{Quasi-periodic oscillations in superfluid magnetars}

\author[A. Passamonti $\&$ S. K. Lander]
{A. Passamonti$^{1}$\thanks{E-mail: andrea.passamonti@oa-roma.inaf.it} , S. K. Lander$^{2}$\thanks{E-mail: samuel.lander@uni-tuebingen.de}
\\ \\  
$^1$ INAF - Osservatorio Astronomico di Roma, Via Frascati 33, 00044 Rome, Italy \\
$^2$ Theoretical Astrophysics, University of T\"{u}bingen, Auf der
Morgenstelle 10, D-72076  T\"{u}bingen, Germany}

\begin{document}

\date{\today}

\pagerange{\pageref{firstpage}--\pageref{lastpage}} \pubyear{}

\maketitle

\label{firstpage}


\begin{abstract}

We study the time-evolution of axisymmetric oscillations of superfluid
magnetars  with  a poloidal magnetic field and an elastic crust, 
  working in Newtonian gravity.
Extending earlier models, we study the effects of composition gradients and entrainment on the magneto-elastic wave spectrum and on the potential 
identification of the observed quasi periodic oscillations (QPOs). We use two-fluid polytropic equations of state to construct our stellar models, which 
mimic realistic composition gradient configurations. The basic features of the axial axisymmetric spectrum of normal fluid stars are reproduced by our results 
and in addition we find several magneto-elastic waves 
with a  mixed character.  In the core, these oscillations mimic the shear mode pattern of the crust   
as a result of the strong dynamical coupling between these two regions. 
Incorporating the most recent entrainment configurations in our models,  we find that they have a double effect on the spectrum: the magnetic oscillations 
 of the core have a frequency enhancement, while the mixed
 magneto-elastic waves originating in the crust are moved towards the
 frequencies of the single-fluid case.
The distribution of lower-frequency magneto-elastic oscillations for our models is
qualitatively similar to the observed magnetar QPOs with $\nu < 155
$Hz. 
 \emph{In particular, some of these QPOs could represent mixed magneto-elastic oscillations with frequencies not greatly different from the crustal modes of an unmagnetised star.} 
We find that many QPOs could even be accounted for using a model with
a relatively weak polar field of $B_{p} \simeq 3\cdot 10^{14}$G,
because of the superfluid enhancement of magnetic oscillations.
Finally, we discuss the possible identification of  625 and 1837~Hz
QPOs either with non-axisymmetric modes or with high-frequency axisymmetric QPOs excited by crustal mode overtones.

\end{abstract}


\section{Introduction} \label{sec:intro}

Magnetars are a class of highly magnetised, slowly rotating neutron
star. In contrast to typical rotation-powered pulsars, a magnetar's
magnetic field is believed to provide the energy reservoir needed to
explain its quiescent activity and flaring episodes \citep{1996ApJ...473..322T}. Most
spectacularly, three of these stars have been seen to 
suffer giant flares, with energy output up to around $10^{47}$
erg. In the aftermath of the more recent two observations, the data
have been good enough to reveal the presence of quasi-periodic
oscillations (QPOs) in the X-ray tail following the initial
outburst. These QPOs span a rather wide range of frequencies, from
tens of Hz up to kHz, and many are only detected in short parts of the
tail, which itself lasts for hundreds of seconds
\citep{2005ApJ...628L..53I,2005ApJ...632L.111S}.


These observations appear to represent the first direct detection of
neutron star oscillations and hence should contain information about the
star's equation of state (EoS), internal magnetic field configuration and 
so on. The problem is disentangling the contributions of these different
aspects of the star's physics given the limited available data; a
flurry of papers since the original detections have aimed 
to do just that, through theoretical modelling of magnetar
oscillations. Initially several studies focused on either the purely crustal oscillation problem~\citep[see e.g.][]{1998ApJ...498L..45D, 2007MNRAS..374} 
or the magnetic mode interpretation by using different models and techniques~\citep[see e.g.][]{2007MNRAS.377..159L, 2008MNRAS.385L...5S,2009MNRAS.397.1607C, 2009MNRAS.396.1441C}. A relevant result found by~\citet{2007MNRAS.377..159L}  and confirmed by other studies was that the spectrum 
of torsional Alfv\'en oscillations in a star with a purely poloidal
field is not discrete but consists of bands of continua. Magnetised
models with a crust show a strong interaction between the magnetic and
elastic waves~\citep{ 2006MNRAS.371..L74,2006MNRAS.368..35L}, which 
form a single family of hybrid magneto-elastic oscillations (for a more detailed description of the literature see Sec.~\ref{sec:QPO}).

With general agreement about the oscillations of a single-fluid
magnetar, attention has returned to the effect of neutron
superfluidity on Alfv\'en oscillations. Earlier qualitative studies
predicted that superfluidity would serve to increase the frequencies
of magnetar QPOs with respect to those of a single-fluid model
\citep{2008MNRAS.391..283V,2009MNRAS.396..894A}. We confirmed this through numerical
simulations, focusing on non-axisymmetric modes in a magnetar core 
\citep{2013MNRAS.429..767P}. Our results suggested that the
higher-frequency observed QPOs could be interpreted as Alfv\'en core 
modes, and we conjectured that matching our core model to an elastic crust could
allow for a set of low-frequency magneto-elastic modes too, thus
accounting for the wide range of frequencies observed. This was very recently confirmed by
\citet{2013arXiv1304.3566G}, in the simplest case where the neutron
fluid is totally decoupled from the charged (proton plus electron)
fluid in the core.  This study found constant-phase oscillations,
  indicating that the QPO spectrum of a superfluid star can become effectively
  discrete at typical magnetar field strengths $B_{\p} \sim 10^{15}$G, where $B_{\p}$ is the magnetic field 
at the pole. 

The neutron superfluid in a neutron star is likely to be
coupled to the proton fluid to some degree, through entrainment \citep{1989tns..conf..457S}. For
strong entrainment, in fact, the dynamics of the 
system would return to that of a single-fluid model. Another important
aspect of multifluid physics in a neutron star is the non-constant
ratio of protons to neutrons; protons make up a smaller fraction of
the outermost part of the core than they do deeper in the star. It is
therefore important to check what influence these effects have on the
star's oscillation spectrum.

Our aim in this paper is to include these two pieces of physics and hence make a
fuller study of superfluid effects on magnetar QPOs. This work focuses 
 on axisymmetric oscillations and is intended to be
complementary to our earlier work on non-axisymmetric magnetar QPOs
\citep{2013MNRAS.429..767P}. We begin by
discussing our unperturbed background models, followed by the perturbation
equations for a magnetised multifluid star with an elastic crust and
the time-evolution code we use to solve them. Next, 
we confirm that our code reproduces earlier results for magnetar QPOs
in the single-fluid limit, before moving on to studying two
representative models for superfluid magnetars. Finally, we compare our
results to the observations and discuss key outstanding issues in
magnetar QPO modelling.

\section{Equations of Motion} \label{sec:Eq}

Three particle species are essential ingredients in a model
neutron star: neutrons, protons and electrons. These appear in
different guises in different regions of the star, depending on the local
density and temperature. In the crust, protons and a fraction of
neutrons are bound in a lattice of nuclei which become progressively heavier
towards the bottom of the crust. When the density exceeds
approximately $4.3 \times 10^{11} \textrm{g cm}^{-1}$ ,  
neutrons may drip out of the nuclei and, if the temperature is below
about $10^{9}$K,  form a gas of superfluid neutrons. Also below this
temperature, the neutrons and protons of the core are expected to
become, respectively, superfluid and superconducting. 

The dynamics of a superfluid neutron star are therefore more complex
than a single-fluid model. However, the number of degrees of freedom
can be reduced, if we note that the electromagnetic force locks
together the various charged particles on very short timescales
compared  
with the typical mode oscillation periods. This means that the charged particles can  then be considered as a single neutral conglomerate of comoving particles.

A basic superfluid neutron star can therefore be  studied from a dynamical point of view  as a two-fluid system. In the core, 
the two constituents are the superfluid neutrons and a neutral mixture
of charged particles that for simplicity we  call `protons'. 
These two-fluid constituents are, respectively, denoted with n and
p. Although the core protons are likely to be superconducting, we
regard them as a normal fluid in this work, where our focus is rather the
effect of neutron superfluidity on magnetar QPOs.
In the inner crust, we may discern a fraction of free superfluid
neutrons which permeates the lattice of nuclei, in which protons and
the remaining part of neutrons are confined. These two 
crust constituents are denoted with f and c for, respectively,  free superfluid neutrons and confined baryons.

Assuming we have small perturbations away from equilibrium, we may
make the usual separation of the full system of equations for
two-fluid magnetohydrodynamics (MHD) into a set governing the stationary background star and a set of time-evolution
equations for the perturbations. We will see that, within our
approximations, the background models are purely fluid with the two
species fully decoupled; entrainment and elastic terms enter only at
the order of the perturbations. We discuss solving the background and
perturbation equations in turn in the next sections.

\begin{figure*}
\begin{center}
\includegraphics[height=74mm]{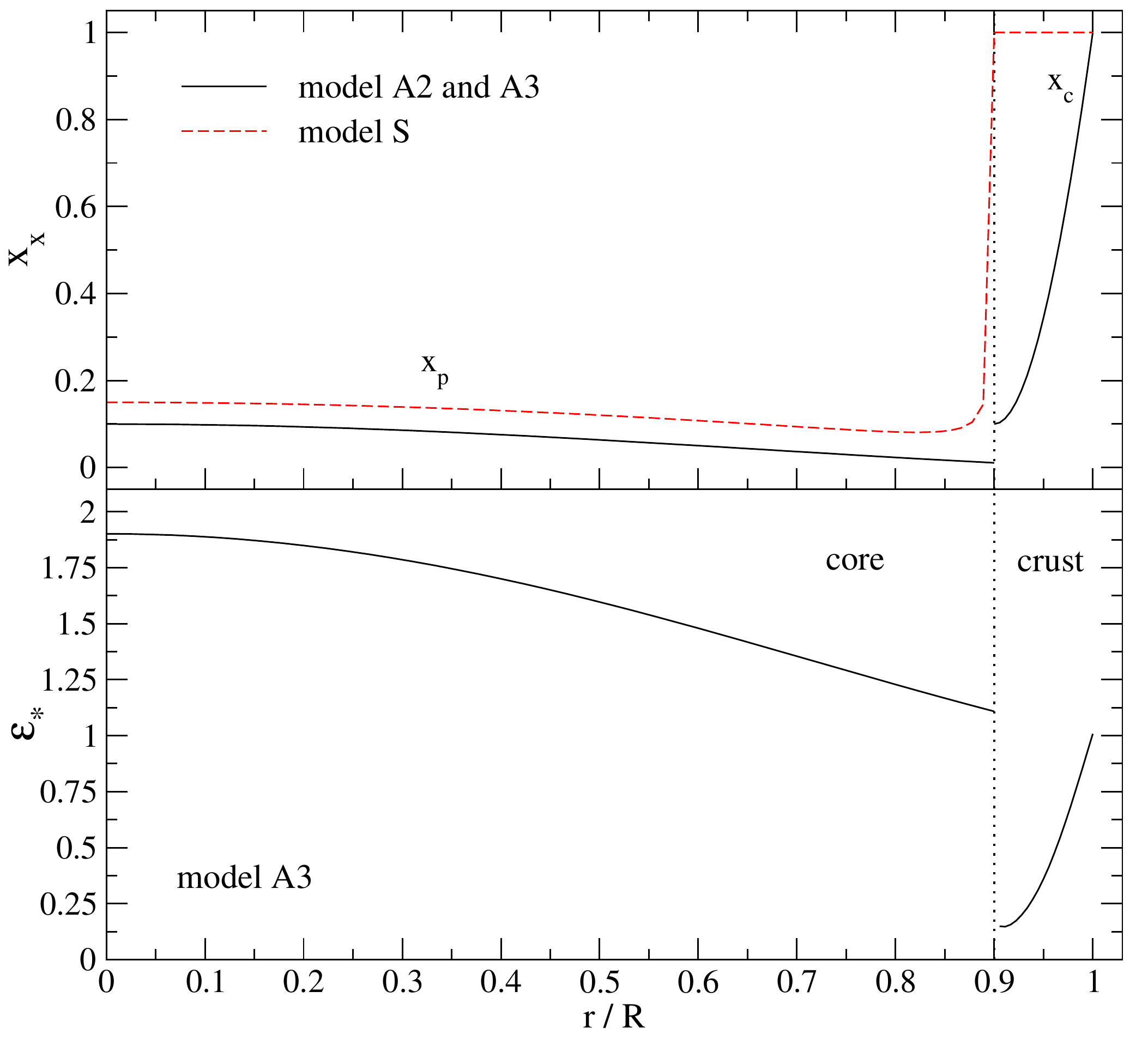} 
\includegraphics[height=80mm]{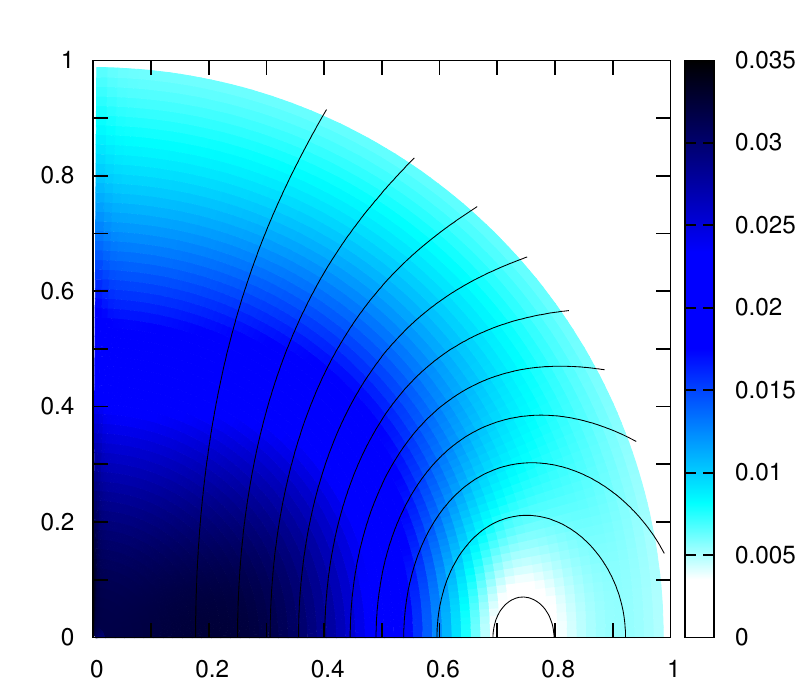} 
\caption{ \label{fig1} 
The upper-left panel shows the proton fraction profile used for models A2 and
A3 (solid black line),  and for model S (dashed red line).
The lower-left panel displays the entrainment parameter $\veps_{\star}$ for model
A3. Note that $\eps_\star=1$ represents a two-fluid limit with
completely decoupled neutral and charged particles, whilst
$\eps_\star=x_{\rm c}$ represents total coupling between them. 
The right-hand panel shows the magnetic field magnitude and direction 
of model S, with a two-fluid core and single-fluid crust.  
The crust/core transition is at $r_{cc} = 0.9R$. The magnetic field configuration 
for the unstratified model A1 is similar, with the closed field line region slightly further
out.}
\end{center}
\end{figure*}

\section{Background models} \label{sec:back_pert}

The background models are constructed as described in
\citet{2012MNRAS.419..732L}. These are 
self-consistent multipolar solutions, accounting for the back-reaction of the
magnetic field on the stellar fluid. The basic effect of neutron
superfluidity, decoupling the protons and increasing the Alfv\'en mode
frequencies, may be accounted for with a basic single-fluid
unstratified background. However, we wish to explore the effect of
more realistic configurations: stratification (i.e. a non-constant
proton fraction), entrainment  and a two-fluid core matched to a single-fluid crust.

We consider stellar models based on a two-fluid polytropic EoS, with
an energy functional $\mathcal{E}$ given by
\begin{equation}
\mathcal{E} = k_{\n} \rho_{\n}^{1+1/N_\n} + k_{\p} \rho_{\p}^{1+1/N_\p} \, , \label{eq:EoS}
\end{equation}
where $k_\x$ are constants and $N_\x$ are the polytropic indices for
each species. This simple EoS allows us to construct models with
composition stratification by choosing different polytropic indices
for protons and neutrons, $N_{\p} \neq N_{\n}$. Our models consist of
a fluid core and an elastic crust, with the crust-core boundary at $r_{cc} = 0.9 R$, where
$R$ is the star's radius. Since we assume that the crust is relaxed for the
background configurations  the equations reduce to those of a
purely fluid model; elastic terms enter only into the perturbation
equations.

In this paper we aim to study oscillations of two `representative' model magnetars,
encapsulating the real physics of the system as best our framework
allows. In addition, we will also need to investigate intermediate cases, in order
to track modes from the well-studied single-fluid regime into new
superfluid territory.

For basic tests, we use an unstratified  model  in which $N_{\p}=N_{\n}=1$, and the proton and confined baryon fraction is 
$x_\p=x_{\crt} = 0.1$. This model has been already used in literature~\citep[see for instance][]{2009MNRAS.396..951P}, 
and we term it `model A1'. 

We generalise this model A1 by adding an ``artificial'' composition gradient both in the 
core and the crust. The relations for $x_\p$ and $x_\n$ try to mimic the profile expected in more realistic models, where the proton fraction decreases 
from the centre to the crust/core transition and the confined baryon fraction increases into the inner crust up to unity at the neutron drip transition. 
This is not fully consistent, but for a pure-poloidal magnetic field geometry the axial modes are decoupled from the polar sector and the only 
background quantity that appears in the dynamical equations is the mass density. Therefore, this is a `controlled' inconsistency that is quite
appropriate for the aims of our current work. 
More specifically, we consider  $x_{\p} = 0.1 \rho $ in the core and $x_{\crt} = 1 + a \rho + b \rho^2$ in the crust, where $a$ and $b$ are two constants 
 that depend on the value of $x_{\crt}$ and its derivative at
 $r_{cc}$.   We call this particular configuration ``model A2'' and its 
 composition stratification is shown  in Fig.~\ref{fig1}.

Another property that characterises superfluid neutron stars is the entrainment, which is a non-dissipative process 
 that couples the dynamics of protons and superfluid neutrons. This effect is due to the strong interaction which 
 limits the free motion of each individual particle species.   The  entrainment can be described, from the dynamical point of view,  in terms of an effective nucleon mass 
 which replaces its own `bare' mass in the dynamical equations \citep{1989tns..conf..457S}. Although our background models are non-rotating, this
quantity affects the dynamics of linear perturbations; see Sec. \ref{sec:perts}.
At the bottom of the inner crust,  the effective mass of superfluid neutrons may be 
 quite large $m_{\n}^{\star} \simeq 14 m_{\n}$, where  $m_{\n}$ is the neutron bare mass. This strong entrainment 
 originates from the neutron Bragg scattering from the nuclei lattice~\citep{2012PhRvC..85c9902C}. In a core with superfluid/superconducting constituents,   
 the neutron effective mass typically is $0.92 \lesssim m_{\n}^{\star}  / m_{\n} \lesssim 1 $~\citep{2008MNRAS.388..737C}.  
 Although we consider normal core protons in our stellar models, we
 investigate also configurations with non-zero core entrainment to
 estimate the effect on the spectrum; note that in this case one
 should include proton superconductivity, so our  core magnetar model is not 
 strictly consistent. 
 We introduce the effects of entrainment 
in our model by using  the approximation to the effective mass profile 
determined by Chamel~\citep{2012MNRAS.419..638P}. We form one of our
two key test models, denoted ``model A3'', by using this entrainment
profile with the same composition gradient described above for model
A2. The entrainment profile is shown in figure \ref{fig1}.

We also study a model with two fluids in the core and only a single
fluid in the crust. This model, which we call `model S', is obtained by choosing $N_{\p}=1.3$
and $N_{\n}=0.6$. In this case we set the central proton fraction
$x_{\p}(r=0) = 0.15$.  The resultant $x_\p$ profile partially mimics
the imposed profile from the previous model, but within a fully
self-consistent framework; this is shown, together with the magnetic
field configuration, in figure \ref{fig1}. This model S can be
thought of as a young magnetar which has
a superfluid core and a normal crust, i.e. 
a star in which the crust's temperature is still above the neutron
critical temperature \citep{2012MNRAS.422.2632H}. This model can be also 
considered an approximation to a star with a strong entrainment in the inner crust, in which the motion of the free superfluid neutrons is almost comoving with 
the confined baryons of the crustal lattice.

We provide our results in dimensionless units which are constructed from the gravitational constant, $G$, the total central mass density, $\rho_0$, and the star's radius, $R$. 
The dimensionless masses for models A and S are, respectively, $M=1.273 \rho_0 R^3$ and $M=1.391 \rho_0 R^3$. 
The crust/core transition is at $r_{cc} = 0.9 R$ where the
  dimensionless mass density is $\rho_{cc} = 0.109 \rho_0$ and
  $0.009 \rho_0$  for models A and S, respectively. The
  differences in the values of $\rho_{cc}$ reflect the fact that our
  toy double-polytrope models cannot reproduce all features of a
  tabulated EoS.

\section{Perturbations} \label{sec:perts}

Let us now derive the evolution equations of axisymmetric ($m=0$) perturbations, where $m$ is the azimuthal number. 
For a purely poloidal or toroidal magnetic field, axisymmetric oscillations can be decoupled in polar and axial perturbations. 
The polar sector describes compressible oscillations which are driven by the radial and theta velocity components and the density perturbation. 
The axial perturbations instead describe an incompressible
motion along  the azimuthal coordinate, which can be completely
characterised for each species by the single velocity component $\delta v^{\phi}_{\x}$, where the index `x' denotes the 
fluid constituents: p and n in the core, c and f in the crust. 

In this work, we focus on the axial sector, which is considered the
most relevant for magnetar QPOs: the oscillation frequencies appear
better matched to the observations (at least in the single-fluid case)
and, being incompressible, the axial modes can be excited   
with a smaller energy \citep{1998ApJ...498L..45D}. However, polar perturbations should not be
completely neglected in a more general magnetar model.  In fact, in
stars with a general mixed poloidal-toroidal magnetic field, the axial
and polar sectors couple and the mode spectrum can be considerably
affected \citep{2012MNRAS.423..811C}. In addition, polar perturbations might be relevant for
understanding the QPOs observed at higher 
frequencies. We intend to study these modes in a future paper.

In a superfluid and magnetised neutron star,  the dynamics of axial oscillations can be described by momentum conservation equations for each fluid 
constituent and by the induction equation~\citep*{2011MNRAS.410..805G}. The mass conservation and Poisson equations are automatically satisfied due to  
the incompressibility of the axial perturbations,  $\Delta  \rho =
\delta \rho = 0$, where $\Delta$ and $\delta$ denote, respectively, the Lagrangian and Eulerian perturbations.
 Since the axial oscillations of a non-rotating
superfluid star with a purely poloidal magnetic field are
characterised by the $\phi$ component
of the velocity and magnetic field perturbations, we omit the coordinate index from all perturbation variables. Therefore $\delta v$ and $\delta B$ stand for 
$\delta v^{\phi}$ and $\delta B^{\phi}$, respectively.

The superfluid neutrons of the core in a non-rotating star obey the following equation:
\begin{equation}
\left( 1 - \varepsilon_{\n} \right) \frac{ \partial \delta v_{\n}}{\partial t}  +  \varepsilon_{\n}   \frac{ \partial  \delta v_{\p}}{\partial t} = 0 \,  ,  \label{eq:EulerN} 
\end{equation}
where the parameter $\veps_{\x}$ accounts for  the entrainment between the nucleons. As already pointed out in Sec.~\ref{sec:back_pert} the entrainment describes the 
relative dragging that a fluid constituent induces on the other component.  This entrainment parameter is related to  the effective mass by $\veps_{\x} = 1- m_{\x}^{\star}  / m_{\x}  $.

To study the dynamics of an elastic crust it is more suitable to introduce the Lagrangian displacement 
 for each fluid constituent \citep{2011MNRAS.416..118A}, 
which is defined in a non-rotating star  by the following expression:
\begin{equation}
\frac{ \partial \xi^{i}_{\x}}{\partial t} =  \delta v^{i}_{\x}  \, .
\end{equation}
Equation~(\ref{eq:EulerN}) therefore can be re-written as follows
\begin{equation}
 \frac{\partial^2 \xi_{\n} }{\partial t^2} =  \frac{\varepsilon_{\n}  }{\varepsilon_{\n}-1}  \frac{\partial^2 \xi_{\p} }{\partial t^2} \, .  \label{eq:EulerNb}  
\end{equation}
The free neutron gas of the inner crust  obeys the same equation~(\ref{eq:EulerNb}) provided we  replace 
the symbols  n with f and p with c.

In the ideal MHD approximation and for non-superconducting protons, the oscillations of the p-fluid in the core are restored by the Lorentz force, and the 
momentum equation reads
\begin{equation}
(1- \varepsilon_{\p} ) \frac{\partial^2 \xi_{\p} }{\partial t^2}  + \varepsilon_{\p}  \frac{\partial^2 \xi_{\n} }{\partial t^2} = \delta \left(  \frac{ f_{L}}{\rho_{\p}}  \right)^{\phi} \, .  \label{eq:EulerP} 
\end{equation}

The confined baryons of the crust interact with the magnetic field via the Lorentz force but can also sustain  shear waves due to the elasticity of the ion lattice. 
Consequently,  the dynamical equation for the confined constituent is given by
\begin{equation}
(1- \varepsilon_{\crt} ) \frac{\partial^2 \xi_{\crt} }{\partial t^2}  + \varepsilon_{\crt}  \frac{\partial^2 \xi_{\f} }{\partial t^2} = \delta \left(  \frac{ f_{L}}{\rho_{\crt}}  \right)^\phi 
+ \frac{1}{\rho_{\crt}} 
\nabla^{j} \sigma^{\phi}_{\, \, j}  \, ,  \label{eq:EulerC} 
\end{equation}
where the elastic stress tensor is defined by
\begin{equation}
  \sigma_{ij} =  \mu \left( \nabla_{i} \xi_{j}^{\crt} + \nabla_{j}
    \xi_{i}^{\crt} \right) - \frac{2}{3}  \mu \left(
    \nabla^{k}\xi_{k}^{\crt} \right) \delta_{ij} \,  ,
\end{equation}
and $\mu$ is the elastic shear modulus. 
For the crust's shear modulus,  we use the approximate relation
\begin{equation}
 \mu = \alpha \rho \,  , 
\end{equation}
which is based on the observation that the specific shear modulus is almost constant through the crust with 
$ \alpha  \simeq 10^{16} \textrm{cm s}^{-1}$~\citep{2001A&A...380..151D}.

Next, by using Amp\`ere's law to replace the electric current, the Lorentz force assumes the following form:
\begin{equation}
f_{i}^{L} = \frac{ B^{j} }{4 \pi } \left( \nabla_{j} B_{i} - \nabla_{i} B_{j} \right) \, , \label{eq:FL}
\end{equation}
while the evolution of the magnetic field is described by the induction equation:
\begin{align}
\partial_{t} \delta B_{i}  = \epsilon^{i j k}  \epsilon_{k l m }  \nabla_{j}  \left(  \delta   v_{\p}^{l} \, B^{m}  \right)  . \label{eq:Beq}
\end{align}
In the crust,  the proton velocity must be replaced with $ \delta v_{\crt}$,  i.e the velocity perturbation of the confined fluid component.

The dynamical problem of the axial axisymmetric oscillations of a superfluid magnetised star with crust can be studied with a single wave equation 
for the proton component in the core and the confined baryons in the crust, as in the single-fluid model. 
Focusing for a moment on the crust, the first step to derive this wave equation is to use equation~(\ref{eq:EulerNb}) for the crust constituents and  remove 
in equation~(\ref{eq:EulerC}) the superfluid neutron variable  $\xi_{\f}$. 
Equation~(\ref{eq:EulerC}) can then be written as follows:
\begin{equation}
 \varepsilon_{\star}^{-1}  \frac{\partial^2 \xi_{\crt} }{\partial t^2}   = \delta \left(  \frac{ f_{L}}{\rho_{\crt}}  \right)^\phi 
+ \frac{1}{\rho_{\crt}} 
\nabla^{j} \sigma^{\phi}_{\, \, j}  \, ,  \label{eq:EulerC2} 
\end{equation}
where $\varepsilon_{\star}$ is a new parameter that accounts for entrainment defined by
\begin{equation}
\varepsilon_{\star} \equiv \frac{ 1-\varepsilon_{\f}}{ 1 - \varepsilon_{\f} - \varepsilon_{\crt}} \, .
\end{equation}
This new entrainment parameter can be written in terms of the effective mass~\citep{2009MNRAS.396..894A}:
\begin{equation}
\varepsilon_{\star} = x_{\crt}  \left(  1 - x_{\f} \frac{ m_{\f} }{ m_{\f}^{\star} } \right) ^{-1} \, .
\end{equation}
The zero entrainment case in which protons and neutrons are entirely decoupled is given by $\varepsilon_\star =  1$.  
The entirely coupled `one-fluid' limit where protons and neutrons comove is given by the $m_{\f} / m_{\f} ^{\star} \to 0$ limit  
which leads to $\varepsilon_\star = x_{\crt} $. 

The second step is to perturb the Lorentz force~(\ref{eq:FL}) and use the time integration of equation~(\ref{eq:Beq}) to replace 
the perturbation of the magnetic field $\delta B$ with the Lagrangian
displacement $\xi_{\crt}$. After some calculation, one may show that the c component obeys the 
following wave equation:
\begin{align}
\rho_{\crt} \veps_{\star}^{-1} \frac{ \partial^2 \xi_{\crt}}{\partial t^2}  & =   A_1 \frac{ \partial^2 \xi_{\crt}}{\partial r^2} 
 + A_2 \frac{ \partial^2 \xi_{\crt}}{\partial \theta^2} 
+  A_3 \frac{ \partial^2 \xi_{\crt}}{\partial r \partial \theta}  \nn \\ 
& +  A_4 \frac{ \partial \xi_{\crt}}{\partial r} 
+ A_5 \frac{ \partial \xi_{\crt}}{\partial \theta} 
+ A_6  \, \xi_{\crt} \label{eq:wv}
\end{align}
where the $A_{k}$ coefficients depends on the background variables and are given in the appendix. 
The core's protons obey the same wave equation  with the coefficients $A_{k}$ taken in the zero shear modulus limit, $\mu = 0$.

\subsection{Boundary conditions} \label{sec:BC}

The regularity condition for equation~(\ref{eq:wv}) and the variable $\xi_{\p}$ at the centre ($r=0$)  lead to 
a zero condition for  the Lagrangian displacement $\xi_{\p}=0$.  
For axial axisymmetric perturbations the variable $\xi_{\p}$ and
$\xi_{\crt}$ also vanish at the magnetic symmetry axis.

At the core/crust interface, we need to impose the continuity of the $\phi$ component of the Lagrangian displacement~\citep{2011MNRAS.416..118A}
\begin{equation}
\xi_{\p} = \xi_{\crt}  \label{eq:cc1} 
\end{equation}
and the continuity of the traction. For a purely poloidal magnetic field and non-rotating star, the $\phi$ 
component of the traction in the crust is given by
\begin{equation} 
\delta t^{\phi} = \left( \mu + B_{r}^{2} \right) \left( \frac{ \partial \xi_{\crt} }{\partial r} - \frac{\xi_{\crt}}{r}  \right) 
+ \frac {B^{r}  B^{\theta} }{r} \left(   \frac{ \partial \xi_{\crt} }{\partial \theta} - \cot \theta \, \xi_{\crt} \right)  \, , \label{eq:tr}
\end{equation}
while in the core it assumes the same form provided we replace the index c with p and set a zero shear modulus ($\mu = 0$). 
From equation~(\ref{eq:cc1}) and the continuity of the magnetic field
across the interface, the traction condition reads
\begin{equation}
B_{r}^{2}  \frac{ \partial \xi_{\p} }{\partial r}   = \left( \mu + B_{r}^{2} \right)  \frac{ \partial \xi_{\crt} }{\partial r}  -   \mu \frac{ \xi_{\crt} }{r}   \, . \label{eq:cc2}
\end{equation}

At the surface of the star our evolutions must satisfy a zero traction condition ($\delta t^{\phi}=0$).

With respect to the equatorial plane, the axial $(m=0)$ perturbations divide in a symmetric and anti-symmetric class. 
The symmetric oscillation modes satisfy $\partial _{\theta} \xi_{\x} = 0$ at $\theta = \pi /2$, while the anti-symmetric modes obey $\xi_\x = 0$.

\subsection{Numerical method} \label{sec:NM}

The wave equation~(\ref{eq:wv}) is evolved in time on a two-dimensional grid which covers  the first quadrant 
of the $\left( r, \theta \right)$ plane, i.e. $ 0 \leq r \leq R $  and $ 0 \leq  \theta \leq \pi/2 $. This restriction to the first quadrant is  feasible 
due to the equatorial reflection symmetries of the axial $m=0$ oscillations. 

The spatial derivatives are discretised by using a second order approximation,  while the time integration of equation~(\ref{eq:wv}) is carried out   by using 
an explicit  iterative Crank-Nicholson method. We use various grid
resolutions to test our results and have also used meshes with finer resolution in the crust 
in order to describe properly the shear waves. However, the results
shown in this paper have been determined with an evenly spaced grid
with resolution $48\times 90$, 
respectively,  in the $\theta$ and $r$ coordinates.  To stabilise the simulations from high-frequency noise,  
we add  an artificial fourth-order Oliger-Kreiss  dissipation term,  $\veps_{\rm D} D_{4} \xi$ , with a small dissipation coefficient $\veps_{\rm D} \sim 10^{-5}$. 

The junction conditions~(\ref{eq:cc1}) and~(\ref{eq:cc2}) at the crust/core interface are used  to determine at any time step the value of $\xi_{\x}$ at $r_{cc}$. 
This is obtained by discretising  equation~(\ref{eq:cc2})  with a side-stencil derivative both in the core and the crust region and then 
using  equation~(\ref{eq:cc1}).

\subsection{Initial conditions} \label{sec:IC}

The initial data of our time evolutions are a linear combination of various shear mode eigenfunctions of an unmagnetised star which are determined 
with an eigenfrequency code~\citep{2012MNRAS.419..638P}. These eigenfunctions are prolonged into the core by  
matching them continuously with polynomial functions in the radial coordinate. The angular part of the initial data must obey the 
boundary conditions described in Sec.~\ref{sec:BC}.

\section{QPO Spectrum} \label{sec:QPO}

In magnetars, the QPO spectrum is believed to originate from  magneto-elastic waves which arise from  the dynamical interaction between the 
global magnetic field oscillations and the shear waves of the crust. The strong magnetic field of magnetars in fact efficiently couples
the dynamics of the core and the crust, and the families of the shear and  Alfv\'en modes are not independent anymore.  It is 
more appropriate to consider them as a unique class of magneto-elastic oscillations.

Many of the observed QPO frequencies fall in the range $\nu < 100$Hz,
suggesting that they may be generated by oscillations with axial parity and symmetry. 
This hypothesis is also supported by the minor energy required to
excite the axial and axisymmetric modes, making them more suitable  for the QPO seismic origin interpretation. 

\begin{table}
\begin{center}
  \caption{\label{tab:1} This table displays the frequencies of the first three axial shear modes determined with the current 
  time domain evolution (TD) and with the frequency domain (FD) approach. The mode frequencies for the $l=2,3,4$ multipoles  
  are shown in dimensionless units $\nu / \sqrt{ G \rho_{0}}$ for an unstratified model with constant component fraction $x_\p=x_\crt = 0.1$ 
   and with no magnetic field.  The results
  obtained with these two different approaches agree to within a few
  percent. }
\begin{tabular}{  c  c c c  c c  c }
  \hline
     &  & TD  & & & FD &   \\  
$ l $  & ${}^l t_{0}$ & ${}^l t_{1}$ & ${}^l t_{2}$ & ${}^l t_{0}$ & ${}^l t_{1}$ & ${}^l t_{2}$   \\
  \hline
  2 & 0.0109  &  0.1920 & 0.3539 & 0.0108  & 0.1906 & 0.3514  \\
  3 & 	0.0177  &	0.1982 & 0.3631 & 0.0170  & 0.1914 & 0.3517  \\
  4 & 	0.0228  &	0.1986 & 0.3638 & 0.0229  & 0.1920 & 0.3520  \\
\hline
\end{tabular}\end{center}
\end{table}

As originally shown by \citet{2007MNRAS.377..159L},  magnetic axial-axisymmetric oscillations form bands of continuous spectra which, 
depending on the magnetic field strength, efficiently absorb the shear modes of the crust whose frequencies lie
inside the continuum. In an ideal system, the Alfv\'en continuum is
generated by the independent motion of each magnetic field line with
its own particular oscillation frequency. Several studies have shown that   long-lived oscillations 
exist at both the turning points and the edges of the continuum \citep{2009MNRAS.397.1607C,2009MNRAS.396.1441C,2011MNRAS.410.1036V}. 
For instance,  in neutron stars with  a purely poloidal magnetic field, 
long-lived oscillations should appear both in the region of open
and closed field lines. In the former sector, two families of QPOs  
have typically been identified for each continuum band, namely  a 
lower  $L_{n}$ oscillation near the last open field line and an upper 
$U_{n}$  wave near the pole. Each class can be further divided into symmetric and antisymmetric oscillations 
with respect to the equatorial plane (see Sec.~\ref{sec:BC}).  Symmetric and anti-symmetric QPOs will be, respectively,  denoted  
with the  $(+)$ and $(-)$ upper indices. 

Another distinct class of QPOs is instead associated with the motion
of the closed magnetic field lines, which we call $C_{n}$ following \citet{2009MNRAS.396.1441C}.   
Differently from the $L_{n}$ and $U_{n}$ families the $C_{n}$ QPOs do not have a preferred symmetry at the equatorial plane,  i.e. they appear 
in the spectrum in both cases with equal frequency. We  omit therefore to specify the  $C_{n}$ oscillation symmetry  in the rest of this work.

Note that the literature on this subject has not been consistent with notation
conventions --- for instance,  in \citet{2012MNRAS.421.2054G} the
lower QPOs are called the edges of the continuum and denoted as
$E_{n}$, while the $C_{n}$  are represented with  $L_n$. 

The inclusion of an elastic crust allows for
shear waves which interact with the magnetic field.  
Within the models described above --- axial axisymmetric oscillations of single-fluid
stars --- shear modes with frequencies inside the continuous spectrum 
are absorbed efficiently by Landau damping within a second or less,
while shear modes that lie outside may persist longer. In any case, the
shear modes become magneto-elastic: their mode frequency and
eigenfunction pattern are both modified by the field~\citep{2011MNRAS.410L..37G, 2012MNRAS.420.3035V}.

In a more general magnetic field geometry, like one with a mixed
poloidal-toroidal field configuration, the toroidal field component may couple oscillations with polar and axial parity, leading to a 
more complex spectrum. In particular, it is expected that the
continuum bands will be destroyed by this coupling and the spectrum
should appear discrete \citep{2012MNRAS.423..811C}.

\subsection{Superfluid physics effects on the spectrum} \label{sf_effects}

Below the superfluid transition temperature $T_{c} \simeq 10^{9}$K,
the dynamics of the superfluid neutrons and protons decouple. These two constituents in general do not comove and as a result  the velocity of the 
Alfv\'en and shear waves now depend only on the proton/confined-baryon mass density: 
\begin{equation}
v_{A} = \frac{ B }{ \sqrt{ 4 \pi \rho_{\crt} } } \, , \qquad  \qquad v_{s} = \sqrt{  \frac{ \mu }{ \rho_{\crt} } } \, . \label{eq:vA}
\end{equation}
Since in the core of a neutron star $x_{\p}  \ll 1$ and  in the crust $  0.1 < x_{\crt}  \leq 1$, the Alfv\'en and
 shear modes of a superfluid system have higher frequencies than a normal matter neutron star.

\begin{table}
\begin{center}
  \caption{\label{tab:2} Frequencies for the first three $l=2$ axial shear modes for an unstratified model with 
   $x_\p = 0.1$, zero magnetic field and with a strong crust entrainment as determined by~\citet{2012PhRvC..85c9902C}. 
   The first column shows the value of the confined baryon fraction in the crust, while TD and FD stand, respectively, for time and frequency domain  
   approach. The time evolutions accurately reproduce  the scaling of the mode frequencies with the entrainment.}
\begin{tabular}{  c  c c c  c c  c }
  \hline
     &  & TD  & & & FD &   \\  
$ x_{\crt} $  & ${}^2 t_{0}$ & ${}^2 t_{1}$ & ${}^2 t_{2}$ & ${}^2 t_{0}$ & ${}^2 t_{1}$ & ${}^2 t_{2}$   \\
  \hline
 0.1	&	0.00347		&	0.0669	&	0.1219	&	0.00368	 &	0.0659	 &     0.1200			\\						
 0.5	&	0.00324		&   	0.0638	&	0.1181	&	0.00356	 &	0.0631	 &     0.1158			\\		
\hline
\end{tabular}
\end{center}
\end{table}

\begin{figure*}
\begin{center}
\includegraphics[height=75mm]{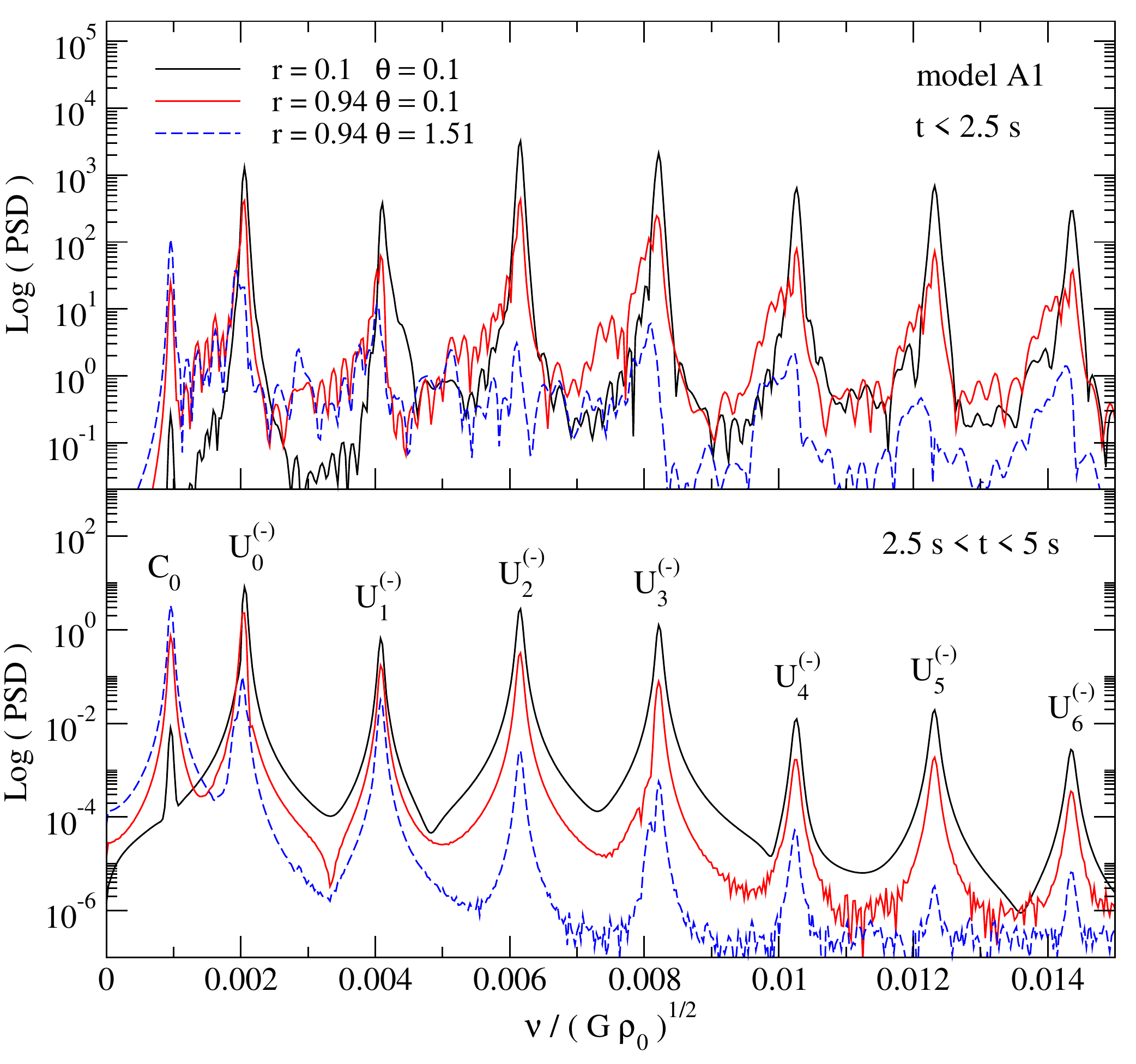} 
\includegraphics[height=75mm]{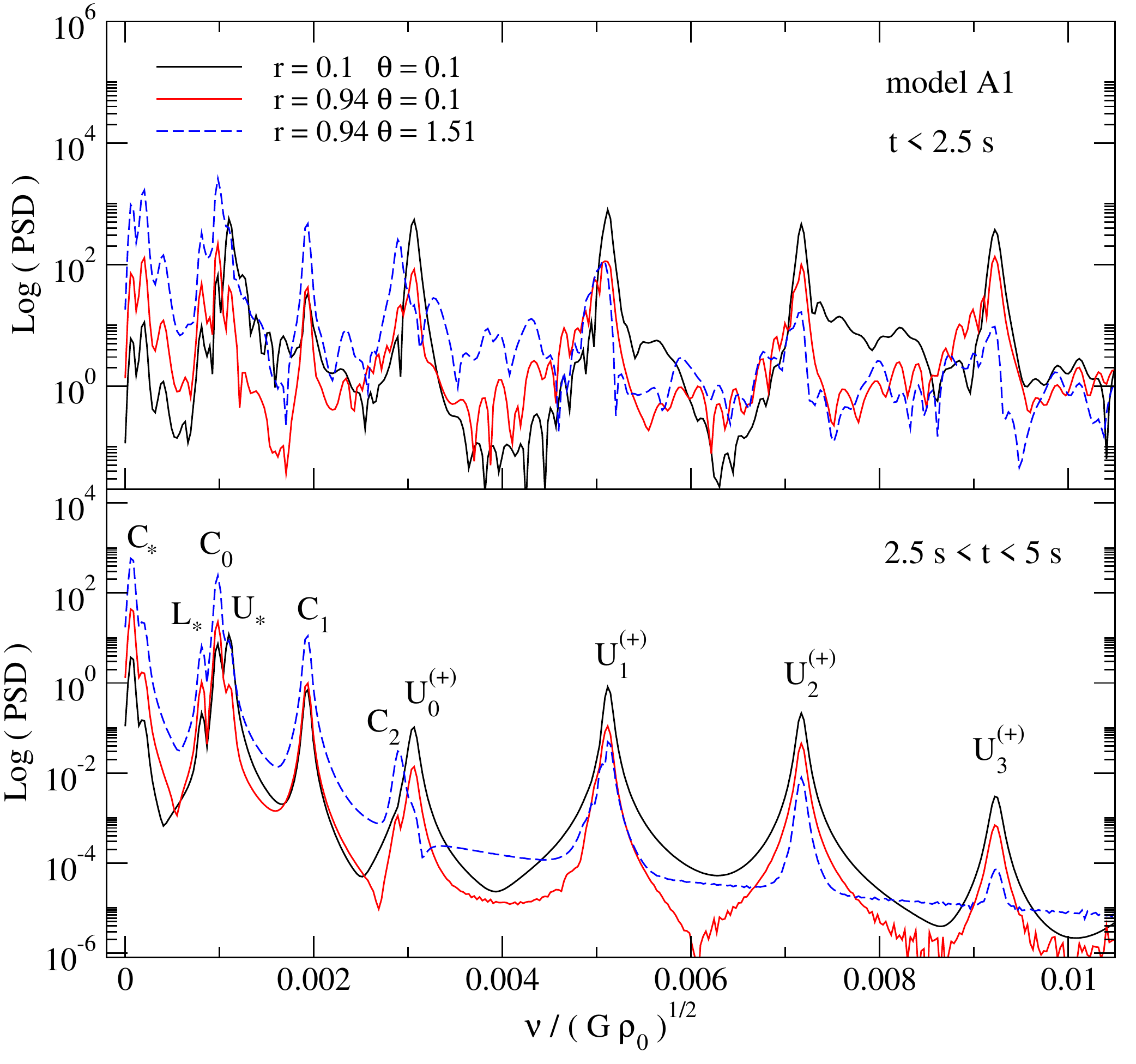} 
\caption{ Spectrum of model A1 with $x_{\p}=x_{\crt}=0.1$ and a poloidal dimensionless magnetic field $B_{p} = 9.48\cdot 10^{-4} \sqrt{G} \rho_0 R$. 
 For a typical neutron star with $M=1.4 M_{\odot}$ and  $R=10 \textrm{km}$ the magnetic field at the pole is $B_{p}=5.4\cdot 10^{14}$G. 
 This figure shows an FFT of the Lagrangian displacement $\xi$ at three different points inside the star and for two periods of the time evolution. 
For the antisymmetric oscillations,  the two left-hand panels display an FFT  at $(r,\theta)=(0.1,0.1)$  (solid black line), $(r, \theta)=(0.94,0.1)$  (solid red line)   
 and $(r,\theta)=(0.94,0.151)$  (dashed-blue line). The two right-hand panels show the same quantities for the symmetric waves. 
 Furthermore, the FFTs performed for a period of $t<2.5 \, \textrm{s} $ are shown in the two upper panels, while the two lower panels display the 
 FFTs for the following time interval $2.5  < t < 5 \, \textrm{s} $.  
 The oscillation frequencies are given in dimensionless units. 
  \label{fig2}}
\end{center}
\end{figure*}

 The entrainment affects the relative motion between the
two fluids and modifies the oscillation frequencies compared with a
completely decoupled two-fluid system. A plane-wave approximation
study~\citep{2009MNRAS.396..894A} showed  that 
the Alfv\'en and shear mode frequencies  in a superfluid star may be described by the following expression:
\begin{equation}
\sigma  = \sqrt{  \frac{ \veps_{\star} }{ x_{\crt} } } \sigma_0 \label{eq:freq} \, , 
\end{equation}
where $\sigma_0$ is the frequency determined in a normal, single fluid star and $\veps_{\star}$ is the entrainment parameter 
introduced in Sec.~\ref{sec:perts}. In the neutron star's core equation~(\ref{eq:freq}) changes with the usual replacement of 
$x_{\crt}$ with $x_{\p}$.
Of course, whilst equation \eqref{eq:freq} provides a qualitative
prediction for the scaling of magnetic and shear modes, which agrees very well 
with our numerical results for simple stellar models with no composition gradients, 
one needs to solve the
perturbation equations to get quantitative results. 

\begin{table*}
\begin{center}
  \caption{\label{tab3} 
 This table displays the frequencies of $C_{n}$ and $U_{n}$  oscillations for
 four stellar models with an averaged magnetic field  $ \langle B \rangle  =10^{15}$G.  This corresponds to 
 a polar field $B_{p} = 3.3 \cdot 10^{14}$G for model S and $B_{p} = 5.4 \cdot 10^{14}$G for model A.   
 The first column provides the stellar model, while the remaining columns shows   
  the oscillation frequencies in dimensionless units $\nu / \sqrt{
    G \rho_{0}} $ and multiplied by $10^{3}$.  }
\begin{tabular}{  c  c c c  c  c c c c c }
  \hline 
 Model   & $C_0$ & $C_1 $ & $U_{\ast}$ & $U_0^{(-)}$ & $U_1^{(-)}$ & $U_2^{(-)}$  & $U_0^{(+)}$ & $U_1^{(+)}$ & $U_2^{(+)}$   \\
  \hline
  S                    &   1.33    & 2.61   & 0.61       & 1.11           &  2.21    & 3.40  & 1.62 & 2.82 & 3.98 \\
  A1  	          &   0.99   & 1.94 & 1.10	 & 	2.06	    & 4.12   & 6.17 &    3.08     &	5.12  &   7.18 \\
  A2                  &  1.86   &  3.70 	& 1.24  &	2.58	    & 5.29	 & 7.69 &     3.91	  &   6.48   &	 9.37 \\
  A3                  &  2.06   &  	$-$              &   1.41         &	   3.33 &	6.72 & 10.19  & 5.01	 &	8.46  &	11.93  \\
\hline
\end{tabular}
\end{center}
\end{table*}

In our previous  paper~\citep{2013MNRAS.429..767P} we performed time evolutions of non-axisymmetric perturbations,
for purely fluid magnetar models. For reasonable values of
$\varepsilon_\star$, the central proton fraction $x_\p(0)$ and the
polytropic indices $N_\n, N_\p$ we found that magnetic modes in a
superfluid magnetar were shifted up in frequency rather dramatically,
by a factor of 6 with respect to a non-superfluid star.

In this work, we add a crust to our superfluid magnetar models,
specialising to axisymmetric oscillations in this case. As well as
allowing for modes restored by elastic forces, the entrainment profile
of our stars will also be seriously affected. In contrast to the
core, where entrainment is relatively weak, recent calculations show that the motion of superfluid neutrons in
the inner crust may be strongly limited by Bragg scattering with the ion lattice. As a result,   
the neutron effective mass in the inner crust can be  $m_{\n}^{\star} \simeq 14 m_{\n}$. 
Due to this strong entrainment,  the dynamics of the two fluids in the
crust approaches that of a single-fluid system. 
As already calculated in non magnetised systems,  the effect of a strong effective mass on the crustal modes is about a $10\%$ correction of the non-superfluid 
case \citep{2009MNRAS.396..894A, 2009CQGra..26o5016S, 2012MNRAS.419..638P, 2013MNRAS.428L..21S}. 
In the next section, we consider also models with this strong entrainment in the crust and study the effects on the 
magneto-elastic waves.

\section{Results} \label{sec:results}

From the evolutions of equation~(\ref{eq:wv}), we determine the spectral properties of axial, axisymmetric oscillations in superfluid models of magnetars. 
We  first test our numerical framework with simplified models and
then study the oscillations of stars with composition gradients and entrainment.

\subsection{Magneto-elastic waves in unstratified stars} \label{sec:modelA}

We study in this section, the oscillation spectrum of a model A1 star, which is a two-fluid polytropic star with crust/core transition located at $r_{cc}=0.9 R$.  
The polytropic indices are $N_{\x}=1$ both in the core and the crust,
which means there is no composition gradient.  
Although it would be possible to choose a different constant composition fraction in the core and crust, we start for simplicity  
with a model where $x_{\p}=x_{\crt} =0.1$, which is qualitatively similar, from the point of view of the axial-axisymmetric spectrum, 
to the single fluid case  ($ x_{\p}=x_{\crt} =1$). The only difference
is that, as expected from equation~(\ref{eq:freq}),   
the oscillation frequencies are shifted to higher values due to the lower proton fraction.  

\begin{figure}
\includegraphics[height=35mm]{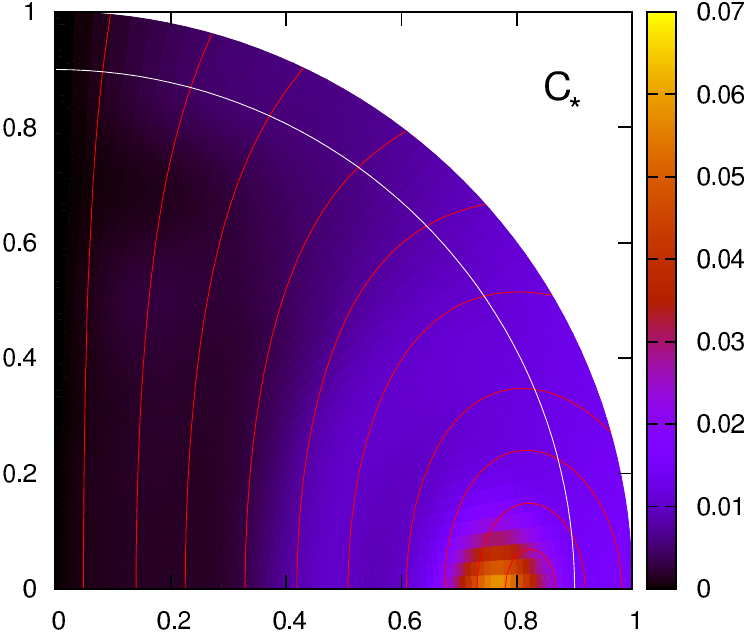}  \quad
\includegraphics[height=35mm]{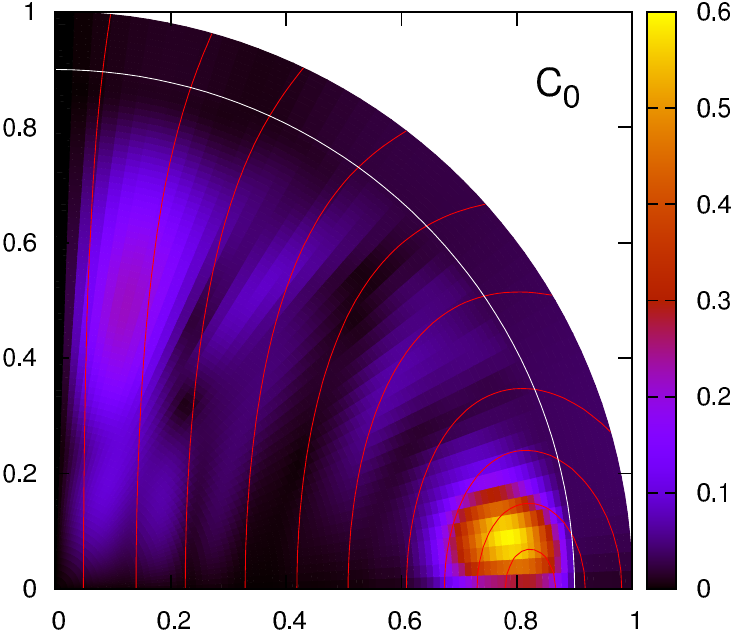} \\
\includegraphics[height=35mm]{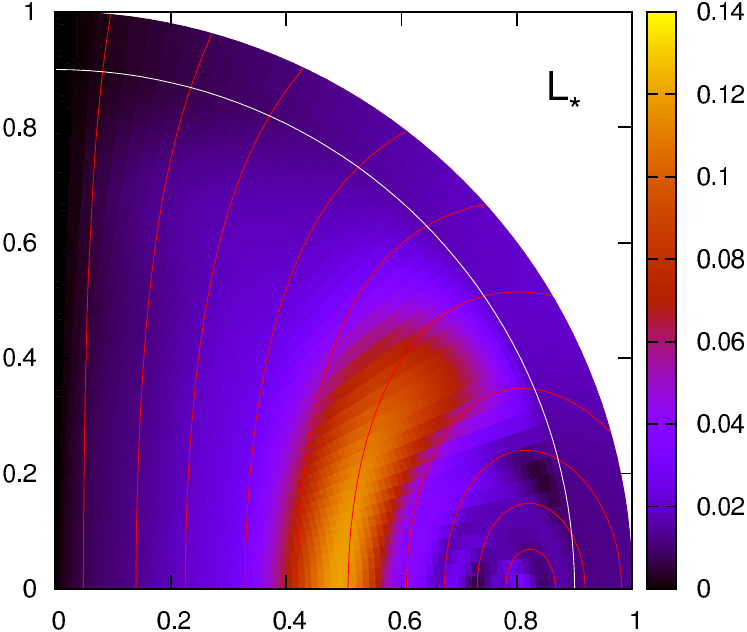}  \quad
\includegraphics[height=35mm]{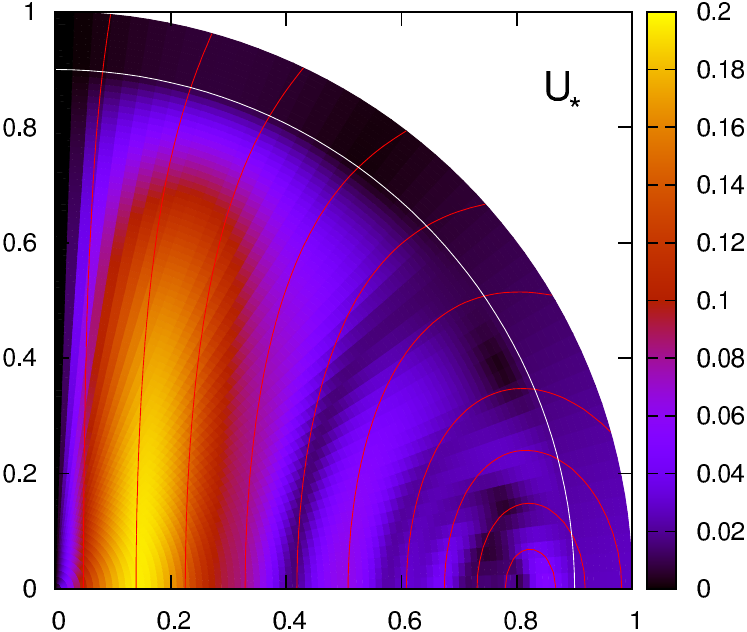} \\
\includegraphics[height=35mm]{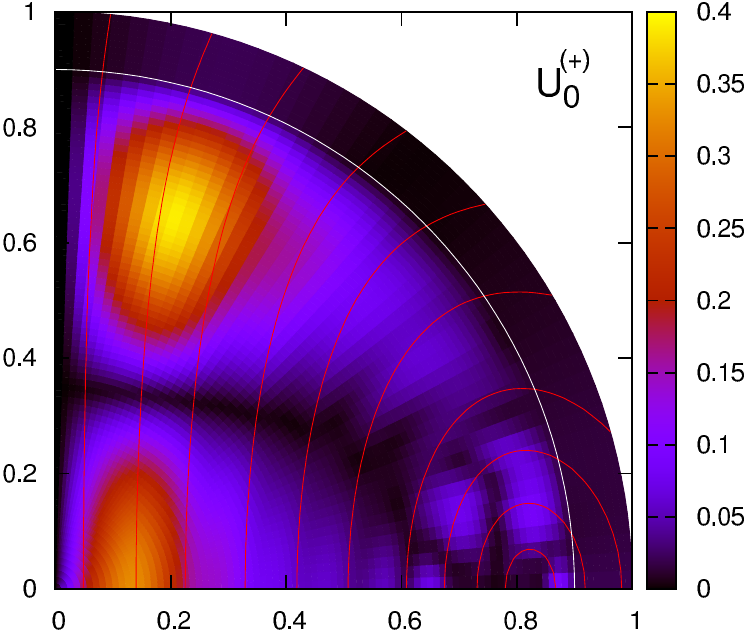} \quad
\includegraphics[height=35mm]{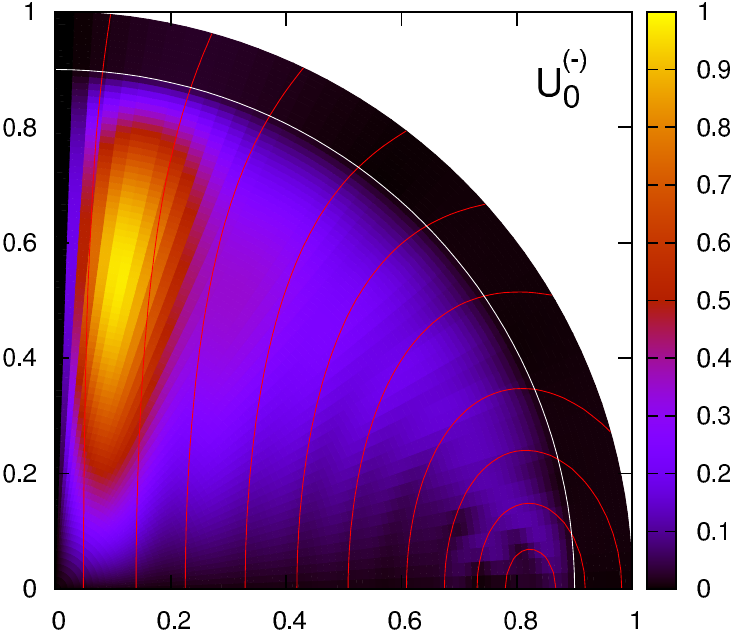} 
\caption{This figure shows the effective 2D-FFT of six magneto-elastic waves for a model A1 star with  magnetic field $B_{p}=5.4\cdot 10^{14}$G. 
From the spectral features of Fig.~\ref{fig2} we select the following modes and show in the  three left-hand panels (from the top): $C_{\ast}, L_{\ast}$ and  $U_{0}^{(+)}$,  
and in the three right-hand panels (from the top):  $ C_{0},  U_{\ast}$  and   $ U_{0}^{(-)}$.   
\label{fig3}}
\end{figure}

To test our numerical simulations, we start with an unmagnetised model and determine the shear modes. We can compare 
the mode frequencies with the results obtained with a numerical code
developed by~\cite{2012MNRAS.419..638P}, which solves the spectrum of
a superfluid neutron star with crust 
as an eigenvalue problem.  The mode frequencies of the 
 fundamental mode and the first two overtones are listed in Table~\ref{tab:1} for the $l=2,3$ and $4$ multipoles. The results of the two approaches 
 agree to better than a few percent. 
 We assess the accuracy of our code also for different values of $x_{\crt}$ and entrainment. For instance,  we provide in Table~\ref{tab:2},  the frequency for 
 two unstratified models, one with $x_\p=x_\crt=0.1$ and the other
 with $x_\p=0.1$ and $x_{\crt}=0.5$, and both having a realistic entrainment profile (see Section~\ref{sec:back_pert}).

As already pointed out in Sec.~\ref{sec:back_pert}, we provide our results in dimensionless units. 
If not otherwise stated we translate any dimensionless quantity to physical values by using a fiducial stellar model 
with typical mass $M=1.4 M_{\odot}$ and radius $R=10\textrm{km}$.

The next step in our preliminary analysis is to add a poloidal magnetic field to the same model A1 with $x_{\p}=x_{\crt} =0.1$. 
We consider  different magnetic field strength and  study the spectrum with fast Fourier transformation (FFT) 
 taken at different positions inside the numerical grid. 
  This is a standard  technique  to recognise the continuum bands of the spectrum 
 and the potential discrete modes. 
 In Fig.~\ref{fig2}, we show an example of an FFT of the Lagrangian displacement for a long evolution of about 5~s 
and  for both symmetric and anti-symmetric equatorial conditions (see Sec.~ \ref{sec:BC}). This star has  a magnetic field at the pole 
 $B_{p} = 9.48\cdot 10^{-4} \sqrt{G} \rho_0 R$  ($B_{p}=5.4\cdot 10^{14}$G for our fiducial model).   
 To trace the energy transfer among the various modes, we perform an FFT in two distinct periods of the simulation before and after $2.5$~s. 
 %
 It is clear from  Fig.~\ref{fig2} that after an initial period where several oscillations are excited the system 
 relaxes in a configuration in which only the $U_{n}$ and $C_{n}$ oscillations survive longer. 
  This behaviour and the analysis  of the FFT in other positions inside the star appear consistent with the presence of bands of continuous spectrum, in which 
 long-lived oscillations can arise at the continuum edges, while  oscillations inside the continuum are quickly damped.

   To  support the identification of the various magneto-elastic waves, we determine also an effective 
 FFT in the whole 2D-numerical grid with a code developed by~\cite{Stergioulas:2003ep}. 
Several $C_{n}$ and $U_{n}$ QPOs are easily identified and in addition we also find the symmetric QPO denoted 
as $U_{\ast}$ by~\citet{2012MNRAS.421.2054G}. 
At low frequencies, we find two symmetric magneto-elastic oscillations, which  
 as far as we know,  have not been reported before in the
 literature. We call them $C_{\ast}$ and $L_{\ast}$ and they are, respectively, associated 
 with the closed and open field line regions. These magneto-elastic waves as well as   $U_{\ast}$ exist due to coupling with the crust.  
 In a model without crust these oscillations would not conserve
 angular momentum \citep{2009MNRAS.397.1607C}.
 The effective 2D-FFT of these QPOs as well as of the first $C_{n}$ and $U_{n}$ oscillations are shown 
 in Fig.~\ref{fig3}. 
 

\begin{figure}
\includegraphics[height=75mm]{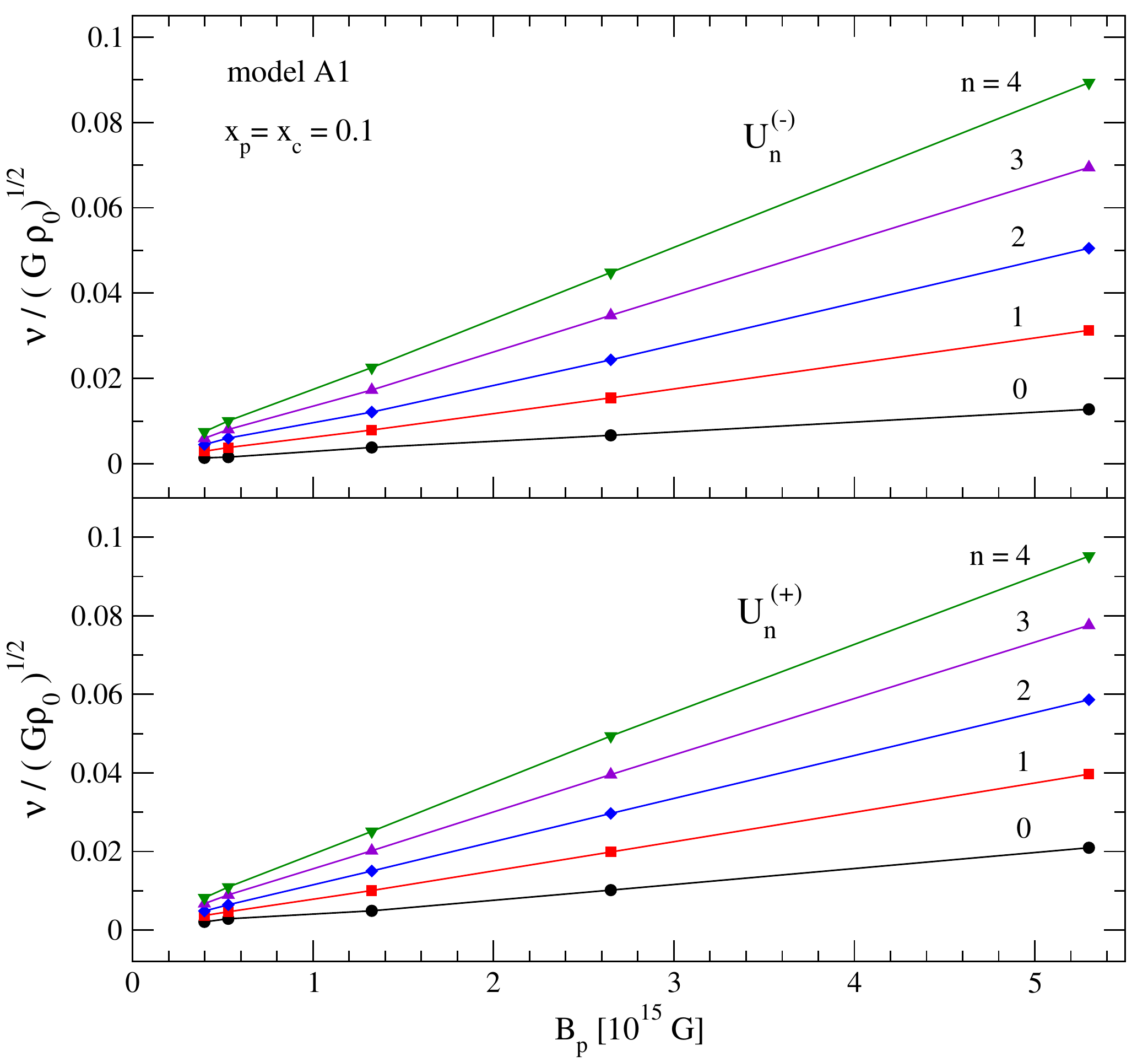}
\caption{ Variation of the $U_{n}$ frequencies  with polar magnetic field $B_{p}$ for a model A1 with $x_\p = x_{\crt} = 0.1$.    
$B_{p}$ is shown on the horizontal axis in physical units for the fiducial model $ M=1.4 M_{\odot}$ and $R=10$km.
The upper and lower panels display, respectively, the antisymmetric and symmetric oscillations. 
\label{fig4} }
\end{figure}

We determine the variation of symmetric and anti-symmetric $U_{n}$ QPOs with the magnetic field and show our results in Fig.~\ref{fig4}. This figure also illustrates 
that we are able to recover the expected scaling in a neutron star with crust, i.e. $U_{n} \simeq \left( 1  +   n \right)  U_{0} $~\citep{2011MNRAS.414.3014C}.

\begin{figure*}
\begin{center}
\includegraphics[height=75mm]{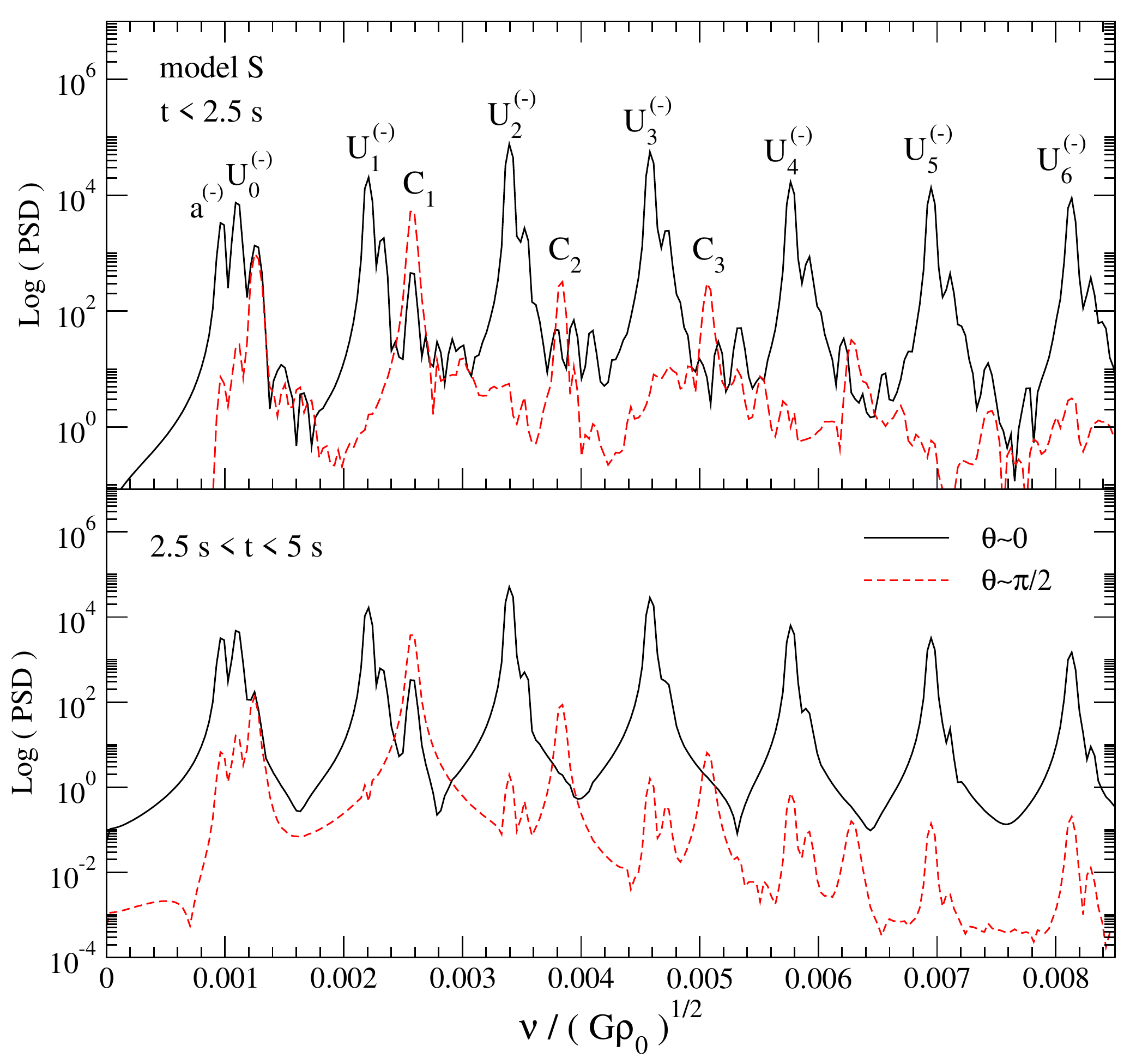} 
\includegraphics[height=75mm]{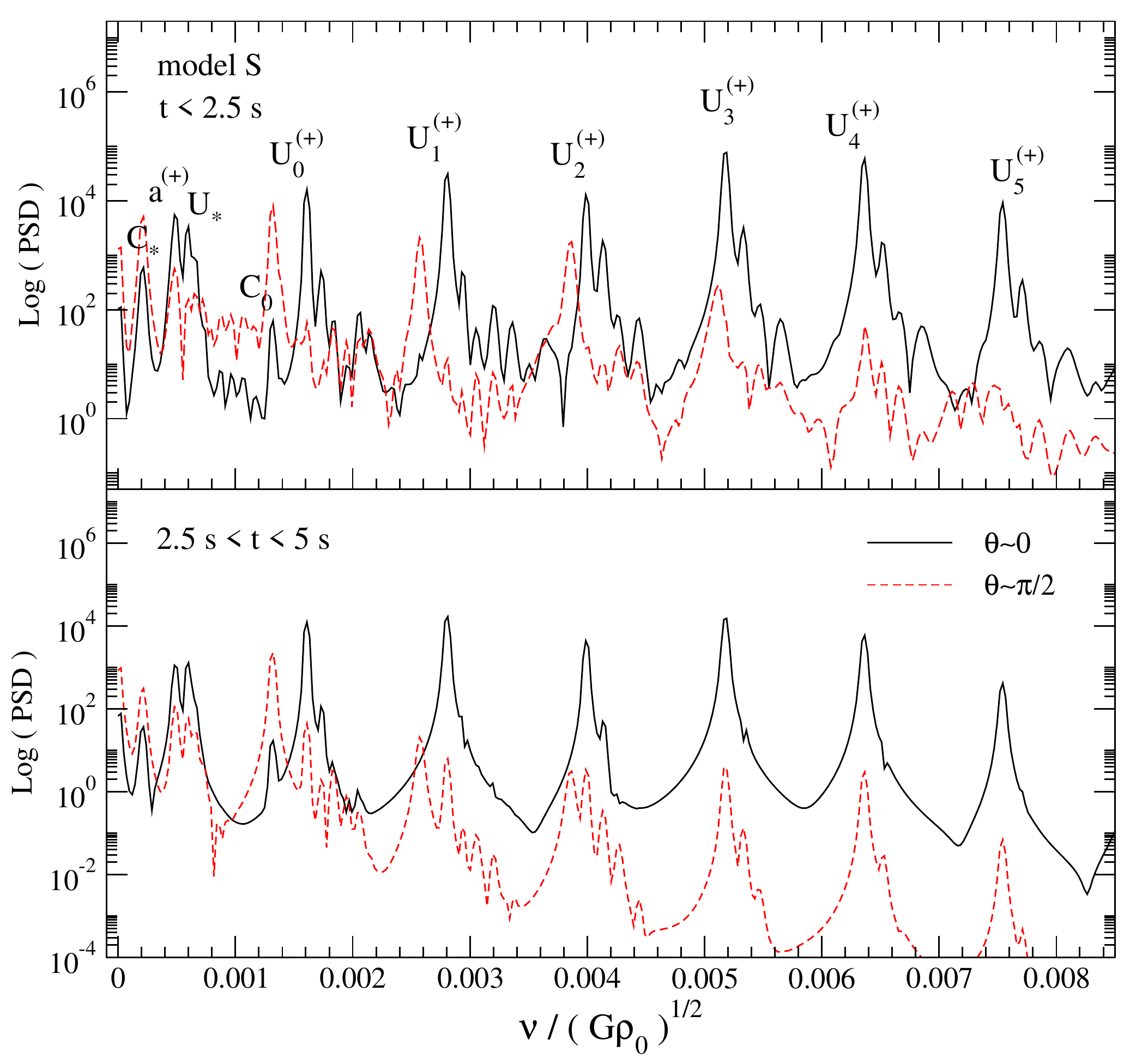} 
\caption{Axial axisymmetric spectrum of magneto-elastic waves for model S, which has  composition gradients in the core. 
The magnetic field at the pole is $B_{p} = 6.4 \cdot 10^{-4}  \sqrt{G} \rho_0 R$. For a typical neutron star with $M=1.4 M_{\odot}$ and 
$R=10 \textrm{km}$ the magnetic field is $B_{p}= 3.3 \cdot 10^{14}$G.  This figure displays the power spectrum of the Lagrangian displacement $\xi$ taken inside the 
crust near the magnetic  axis (solid-black lines) and equator
(dashed-red lines). The two upper panels show  FFTs performed within
2.5~s, while the two lower panels show FFTs taken during $2.5  <
t < 5~\textrm{s}$.  Furthermore, the antisymmetric oscillations are shown in
the left-hand panels and the symmetric ones in the right-hand panels. 
\label{fig5} }
\end{center}
\end{figure*}

 Another method to determine the continuum/discrete nature of the
  spectrum is the study of the mode  phase. In their superfluid
  models, \citet{2013arXiv1304.3566G} find QPOs with constant phase
  when $B_{\p} \sim 10^{15}$G, indicating that these are discrete modes. To investigate this issue, we studied an unstratified 
model A with $x_{\p} = 0.05$ and  $x_{\crt} = 1$ (single-fluid crust), i.e. with a composition fraction similar to that used by~\citet{2013arXiv1304.3566G}, 
and  increased the numerical resolution to $96 \times120$ for, respectively, the $\theta$ and $r$ coordinates.   
Our results of the QPO phase, however, do not show convincing evidence of the continuum/discrete transition. 
This may be due to differences
in the models and framework, but it is difficult to make detailed
comparisons at this stage, given the brevity of the
\citet{2013arXiv1304.3566G} paper.    For the other superfluid models of this work (model S and A3), 
we studied the phase and found the same conclusions as for the unstratified model A.

In models in which  either the component fraction or entrainment assumes the same constant value in the entire stellar volume, the spectrum is expected to shift globally at different frequencies. This is due 
to the same dependence of the Alfv\'en and shear modes on the proton fraction and entrainment (see equations~\ref{eq:vA} and~\ref{eq:freq}). We have also tested the good behaviour of our numerical results by 
varying $x_{\p}=x_{\crt}$ and $\eps_{\star}$,  and found the expected mode frequency scaling. We do not report this test here as 
the scaling is very similar to the non-axisymmetric modes,  which we have already 
considered in~\cite{2013MNRAS.429..767P}. Having confirmed that our
code reproduces earlier results for magneto-elastic waves, we now move
on to consider more realistic configurations.

\subsection{Effects of composition stratification} \label{sec:modelAgrad}

The core and the crust of a neutron star have  composition gradients which depend on the EoS. 
 In general, in the core the proton fraction decreases from the star's centre towards the crust, while 
the fraction of confined baryons increases from the bottom of the crust up to the neutron drip density, 
where all the nucleons are locked in the nuclei and $x_{\crt}=1$.

\begin{figure}
\includegraphics[height=35mm]{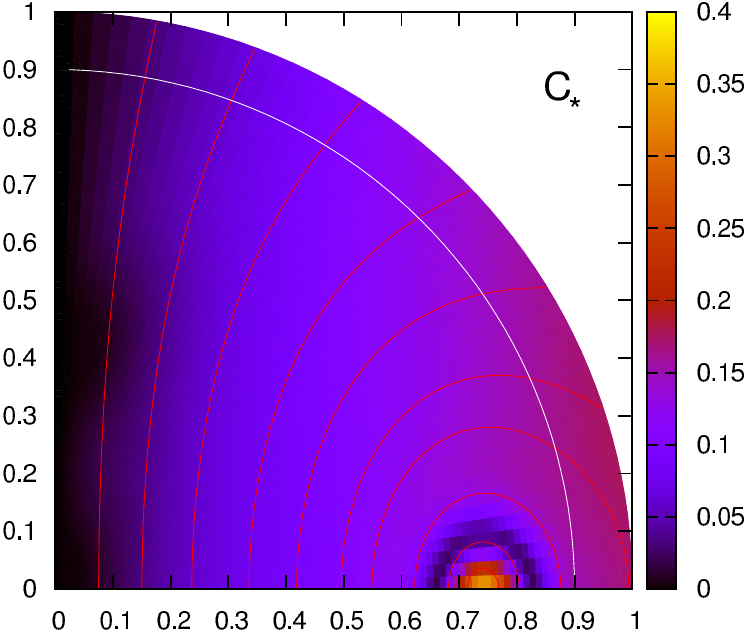}  \quad
\includegraphics[height=35mm]{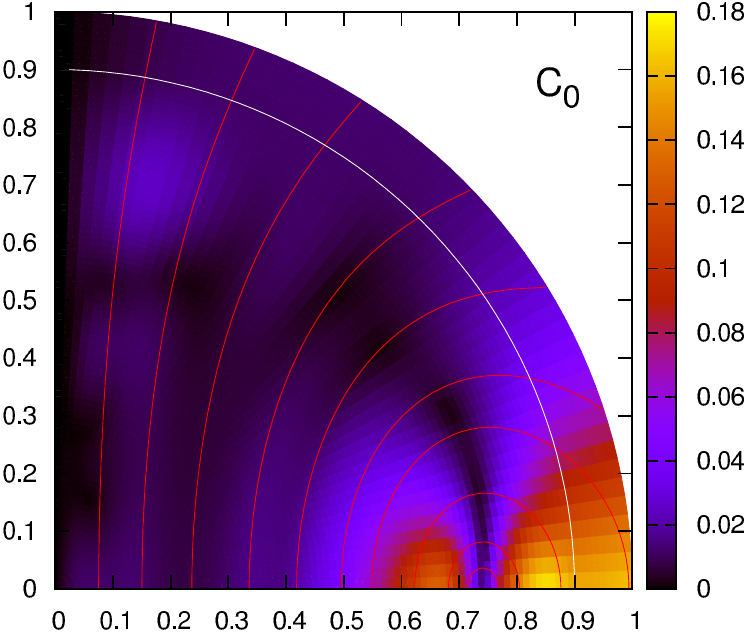}  \\
\includegraphics[height=35mm]{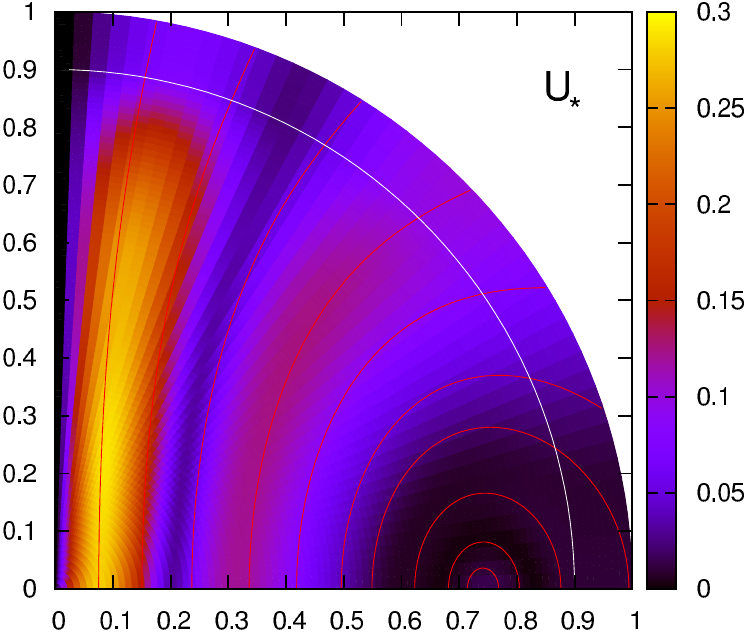} \quad
\includegraphics[height=35mm]{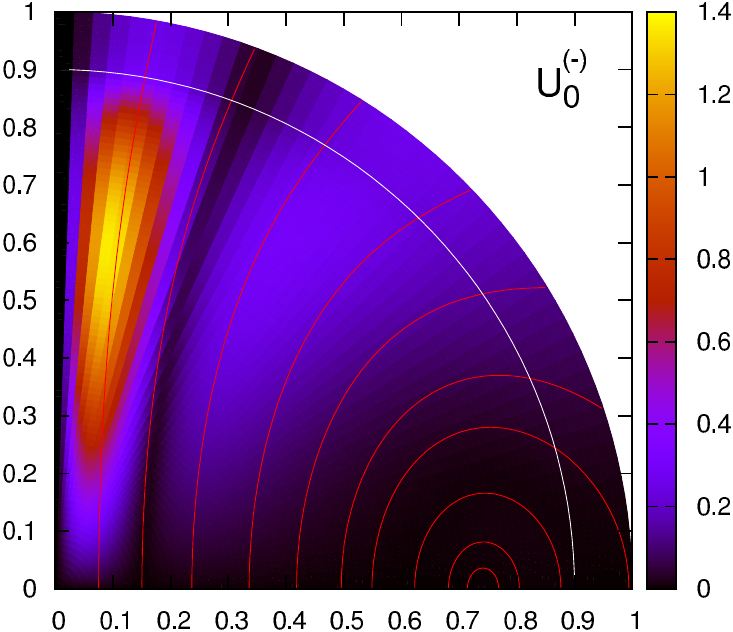} \\
\includegraphics[height=35mm]{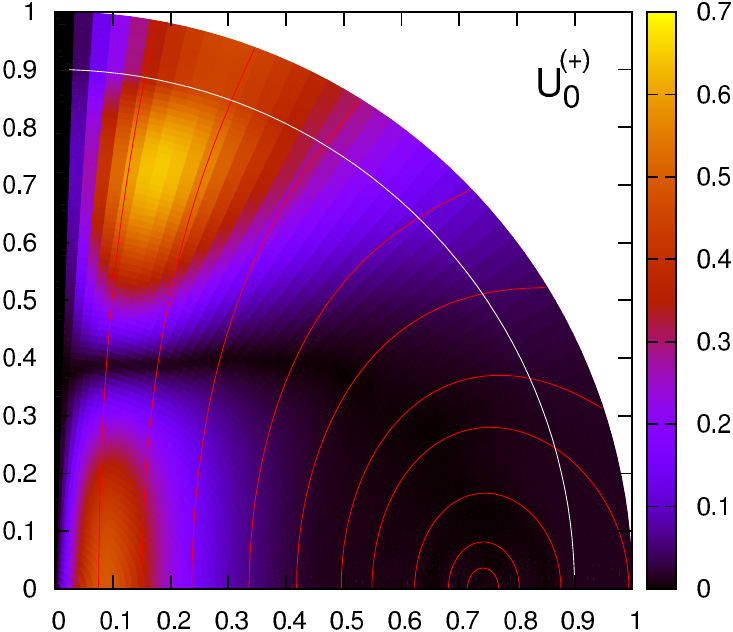}  \quad
\includegraphics[height=35mm]{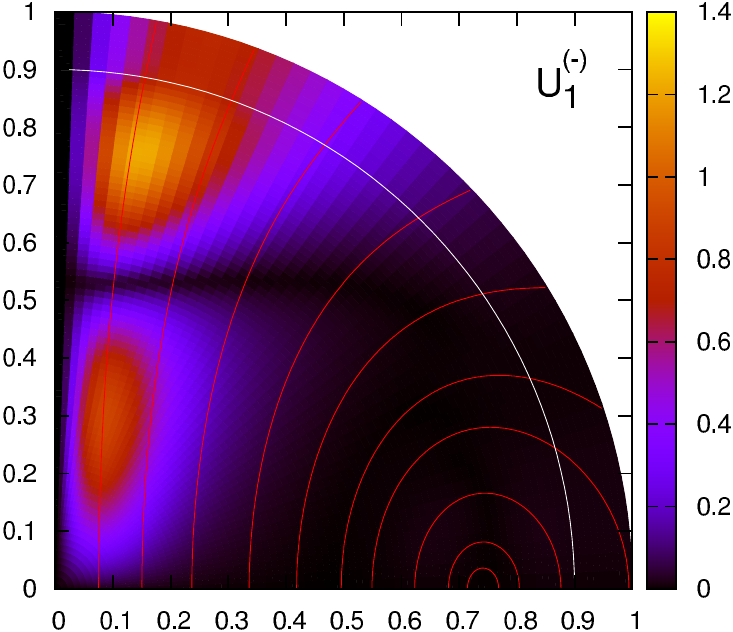} \\
\includegraphics[height=35mm]{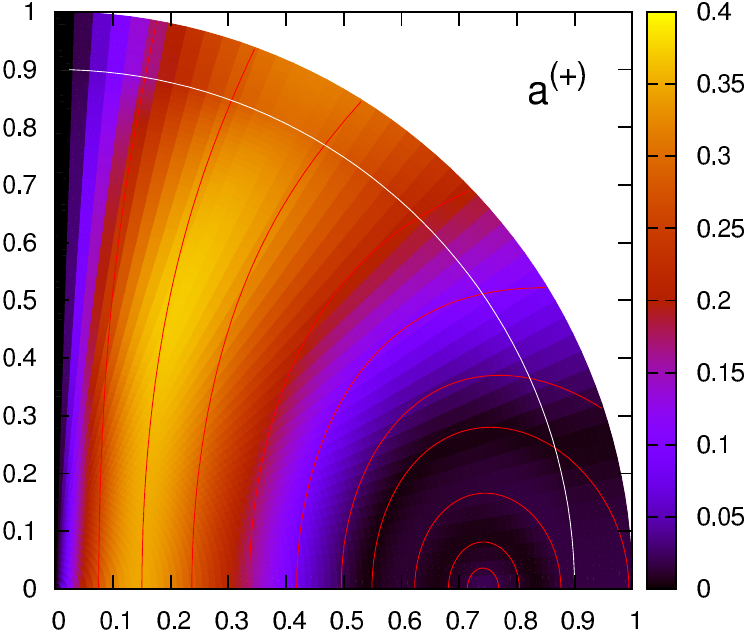} \quad 
\includegraphics[height=35mm]{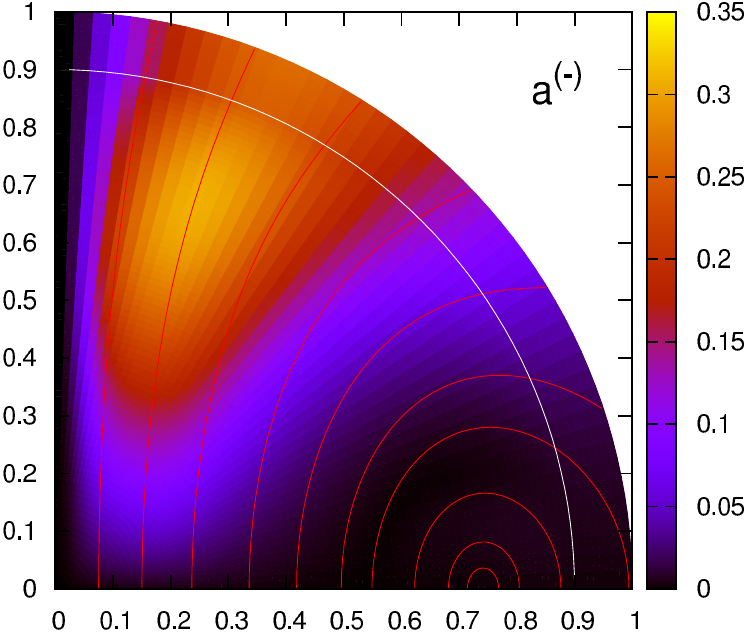}  
\caption{This figure shows the effective 2D-FFT of eight low frequency magneto-elastic waves for model S, whose spectrum is shown 
in Fig.~\ref{fig5}.
The four left-hand panels (from the top) display 
the $C_{\ast}, U_{\ast} , U_{0}^{(+)}$ and $a^{(+)}$ modes, while the  four right-hand panels (from the top) 
show $C_{0} , U_{0}^{(-)} , U_{1}^{(-)}  $ and $a^{(-)} $. The red
curves denote the magnetic field lines and the white curve the crust/core transition.  
\label{fig6}}
\end{figure}

We first study the effects of composition stratification using our
model S. As described in  Sec.~\ref{sec:back_pert}, this is a
two-fluid system in the core which matches a single fluid in the
crust (see Fig.~\ref{fig1}),  as a result it has composition gradients
only in the core. This stratification has been determined by choosing two different polytropic indices for protons and neutrons, respectively, 
$N_{\p} = 1.3 $ and $N_{\n} = 0.6$. 
Considering the initial data introduced in Sec.~\ref{sec:IC}, we provide a linear combination of the first $l=10$ multipoles of the fundamental axial shear modes $^l t_{0}$ 
of a superfluid unmagnetised star.  
These modes reside in the lower frequency band  of the spectrum  and may potentially interact with the global magnetic waves. 
For a star with $B_{p}=3.3\cdot 10^{14}$G and average magnetic field $\Bav = 10^{15}$G, 
we show in Fig.~\ref{fig5} the spectrum of the magneto-elastic waves determined by an FFT performed  inside the crust 
near the magnetic axis and equator. As for model A1, we show the spectrum in two periods of the evolution (before and after 2.5~s) in order to 
identify the long lived oscillations. The properties of the spectrum appear similar to the unstratified model A1, as there is  the same sequence of modes 
albeit at different  frequencies. However, for the same magnetic field strength of model A1, an effective 2D-FFT  of the various modes shows that 
 the magneto-elastic waves in model S also permeate  the crust and can be potentially relevant for the identification of the observed QPOs. 
  This feature 
 is shown in Fig.~\ref{fig6} for some of the magneto-elastic oscillations.  
 The penetration of global oscillations into the crust may be due to the smaller difference between the Alfv\'en 
 and shear velocities at the crust/core interface~\citep{2013arXiv1304.3566G}.
The two lower panels of Fig~\ref{fig6} show the 2D-FFT of  two magneto-elastic waves that we term 
$a^{(+)}$ and $a^{(-)}$, respectively. Their frequencies are near the $U_{\ast}$ and $U_{0}^{(+)}$ QPOs and might be identified in principle 
with the lower QPOs $L_{\ast}$ and $L_{0}^{(+)}$. Their oscillation pattern, however, is not related to the last open field lines and reaches the 
star's surface. Furthermore, their frequencies scale with the magnetic field and not with the shear modulus. For these reasons, we prefer to name them 
differently, as they could be either discrete Alfv\'en modes~\citep{2011MNRAS.414.3014C} or $L_{n}$ oscillations with magnetic modified pattern. 

We have also explored the effect of composition gradients using model
A2. Unlike model S, this also has a gas of superfluid neutrons in the inner crust. 
For the same initial conditions of model S, we consider a star with
$B_{p}=5.4 \cdot 10^{14}$G which corresponds to an average magnetic field $\Bav = 10^{15}$G. Rather similarly to models S and A1, we
found several peaks in the early part of the simulation, including
some of crustal origin. However, these crustal oscillations were quickly absorbed leaving
edges of the continuum as usual. More details on these oscillations are given in Sec.~\ref{sec:entr}
A frequency comparison of models A1, A2 and S may be found in Table~\ref{tab3}.

\subsection{Effects of entrainment} \label{sec:entr}

As described in Section \ref{sf_effects}, entrainment represents a
coupling between the neutral and charged components of a neutron star,
which would otherwise move independently. In this subsection, we
explore the oscillation spectrum of our model A3, which combines the effects of composition stratification and
entrainment, as shown in Fig. \ref{fig1}. It is worth noticing that  $\veps_{\star}$ is larger and smaller than unity, respectively, 
in the core and  the inner crust. Therefore, we expect an  opposite effect on the magnetic- and elastic-dominated modes. The former 
which are located mainly in the core should be shifted at larger frequencies with respect to a zero entrainment model, while the 
shear modes move towards the lower frequency band.  Since $\veps_{\star}$ is not constant we  expect  that 
each mode should  perceive a different value of entrainment which depends on the region where 
the mode  is predominantly located.

\begin{figure*}
\begin{center}
\includegraphics[height=75mm]{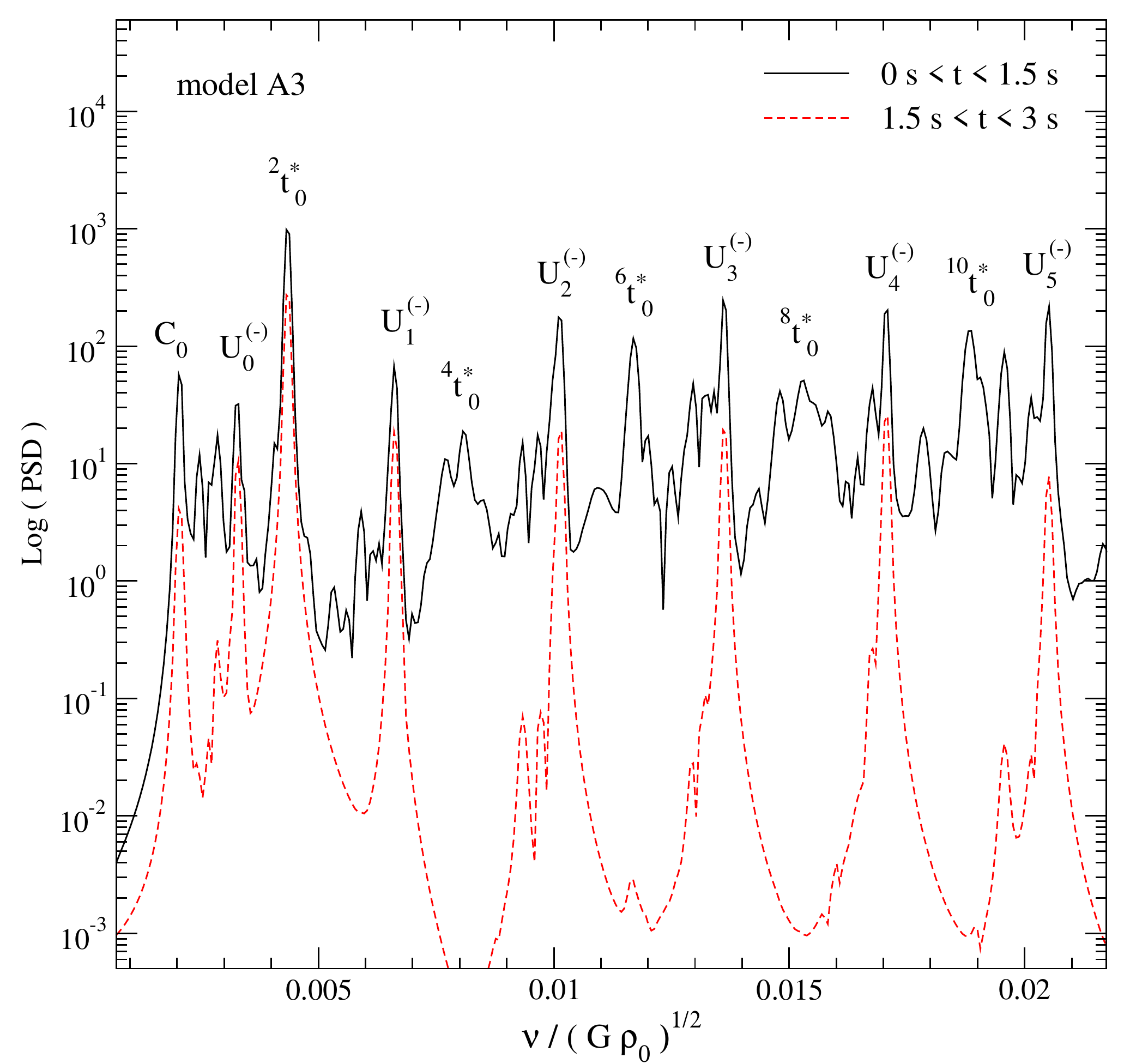}
\includegraphics[height=75mm]{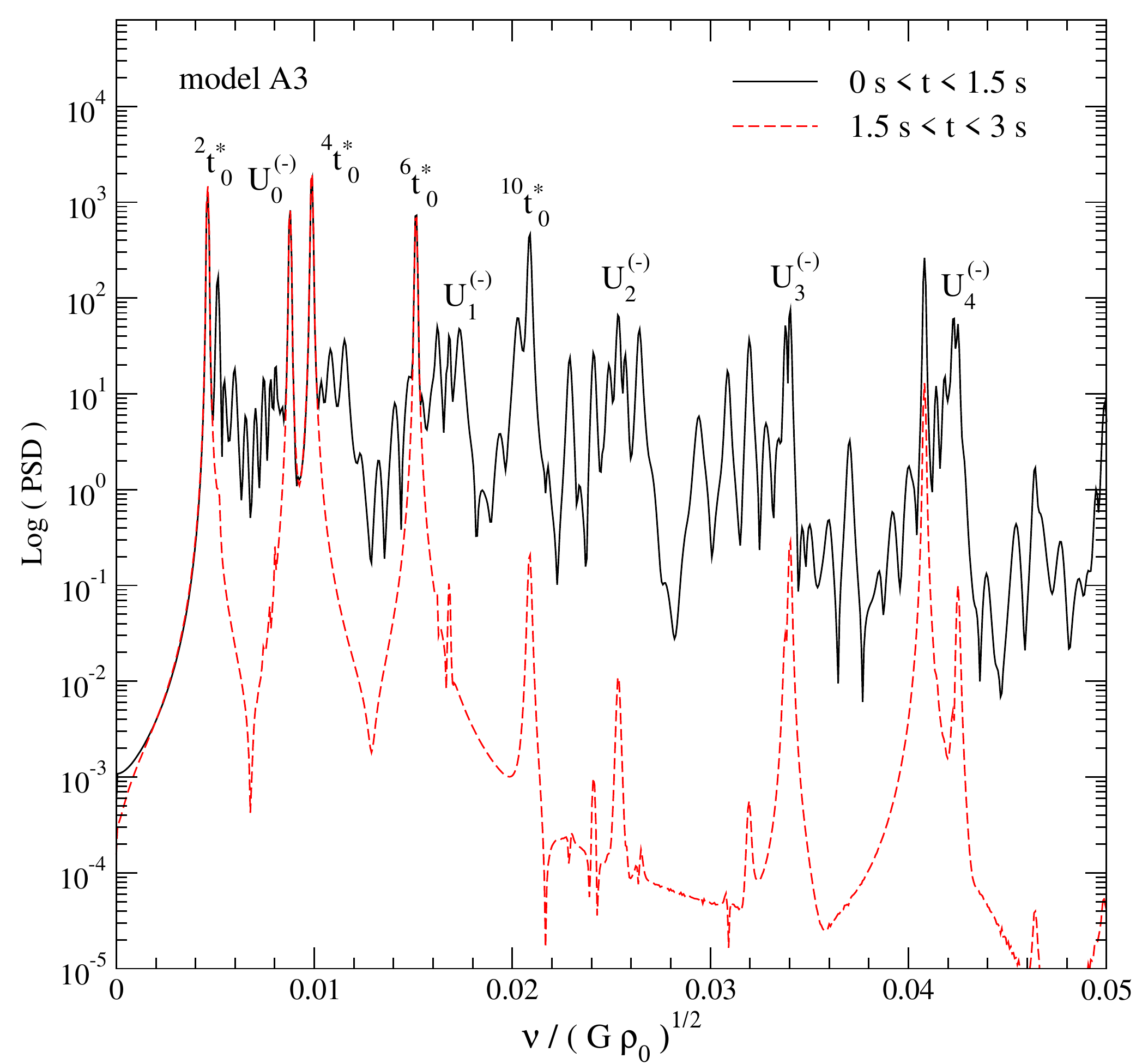}
\caption{Axial axisymmetric spectrum of the magneto-elastic modes for
  model A3, with composition gradients and a strong entrainment in the
  inner crust (Chamel model).  This figure shows the FFT of the
  antisymmetric oscillations (with respect to the equator) extracted inside the star at $r=0.933$ and $\theta=1.505$ for two distinct periods.  The dashed black line denotes the FFT for 
$ t \lesssim 1.5$s, the dot-dashed red line denotes the spectrum for $1.5 \lesssim  t \lesssim 3$s.   
  The left-hand panel displays the spectrum for a  magnetic field  $B_{p}=5.4\cdot 10^{14}$G, while the right-hand panel shows the 
  $B_{p}=1.35 \cdot 10^{15}$G case.   The  physical values for the time and $B$-field are referred to a star with  $M=1.4 M_{\odot}$ and 
$R=10 \textrm{km}$.  
  \label{fig7}}
\end{center}
\end{figure*}

\begin{figure*}
\begin{center}
\includegraphics[height=35mm]{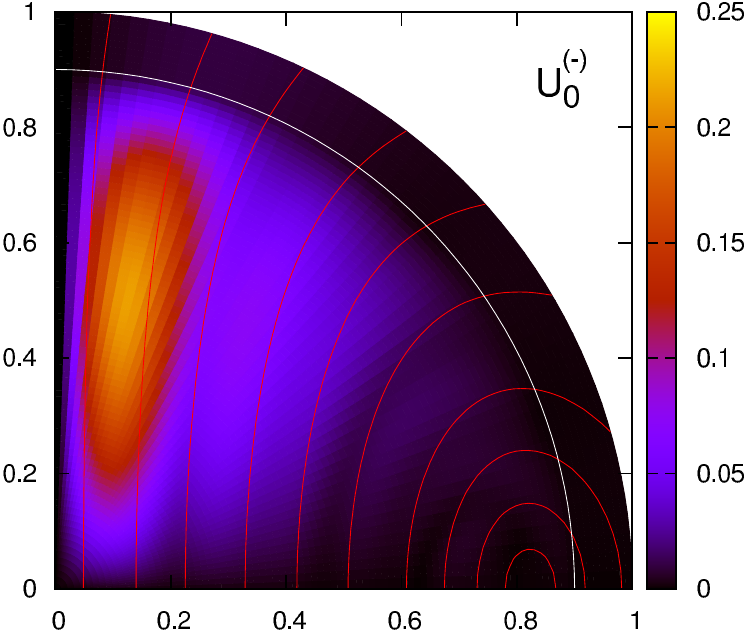} \, 
\includegraphics[height=35mm]{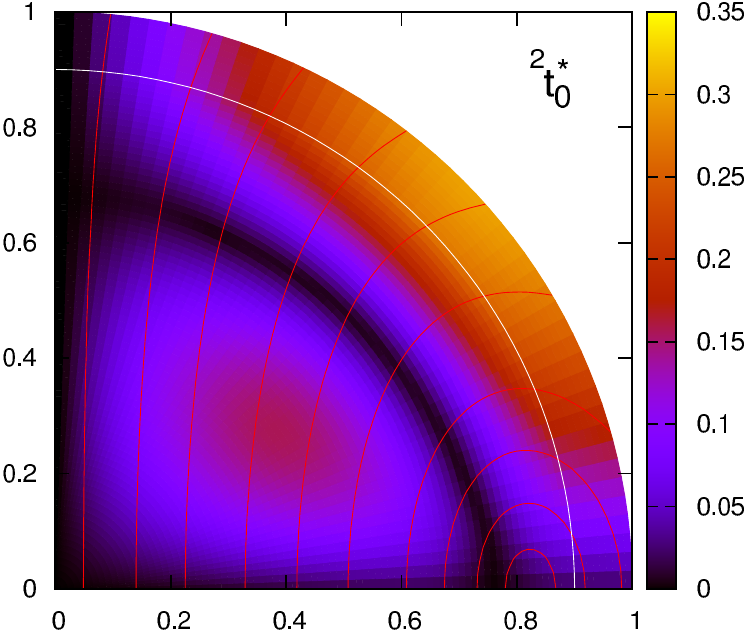}  \,
\includegraphics[height=35mm]{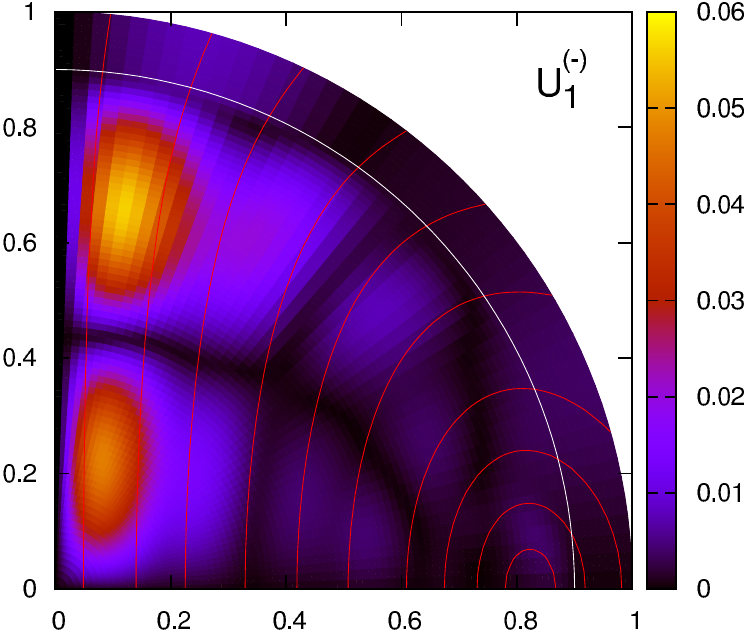}   \,
\includegraphics[height=35mm]{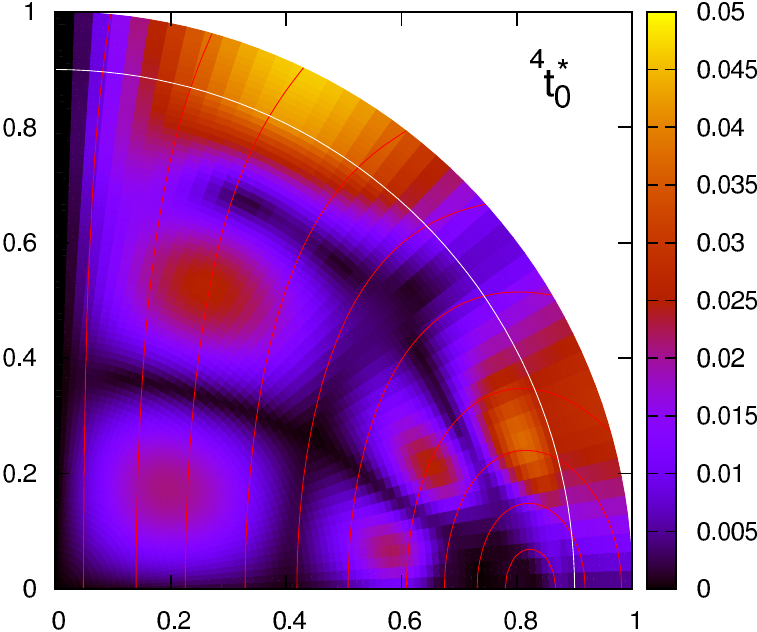}  \\
\includegraphics[height=35mm]{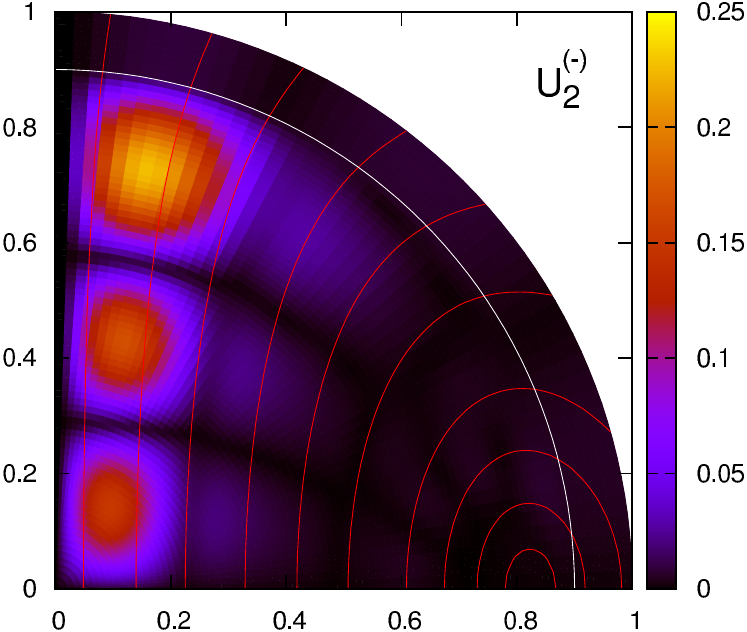}   \, 
\includegraphics[height=35mm]{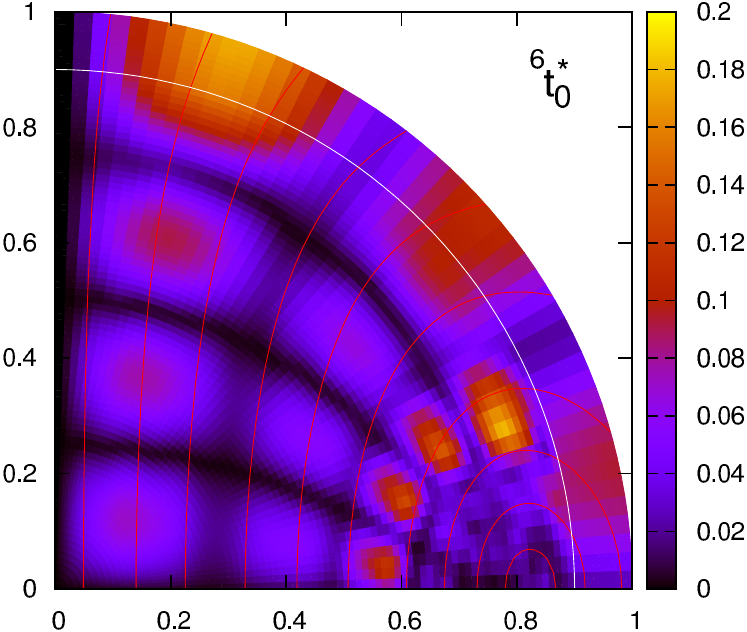} \, 
\includegraphics[height=35mm]{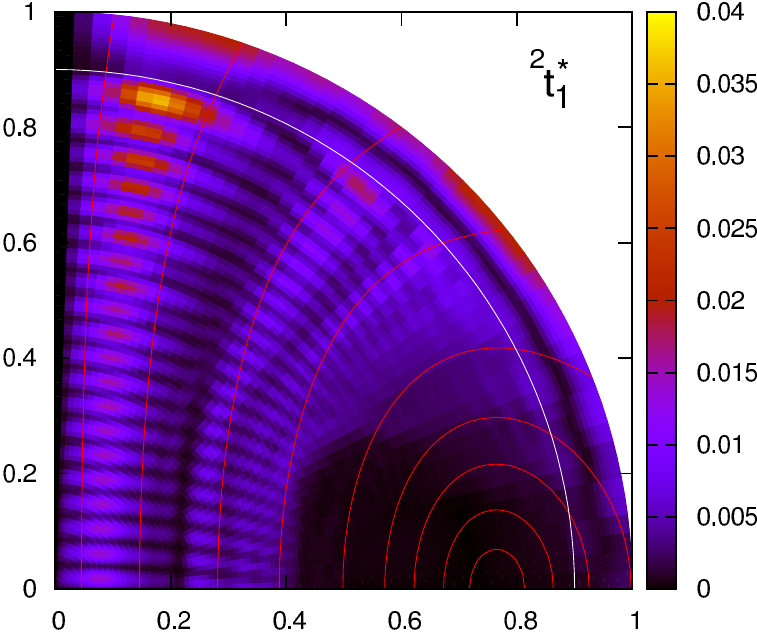}  \, 
\includegraphics[height=35mm]{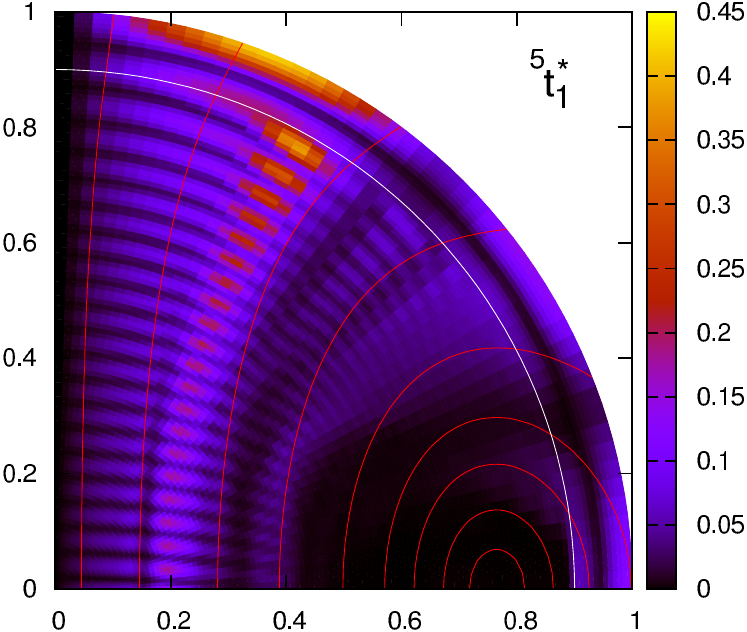} 
\caption{This figure shows the effective 2D-FFT of eight magneto-elastic oscillations for model A3  with 
magnetic field $B_{p}=5.4\cdot 10^{14}$G.  
The four upper panels show (from the left): $U_{0}^{(-)}, ^{2}t_{0}^{\ast}$, $U_{1}^{(-)}$ and $^{4}t_{0}^{\ast}$,  
and the four lower panels show (from the left): $U_{2}^{(-)} , ^{6}t_{0}^{\ast}, ^{2}t_{1}^{\ast}$ and $^{5}t_{1}^{\ast}$. 
\label{fig8}}
\end{center}
\end{figure*}

The results of a long simulation $( t \sim 6\,\textrm{s} )$ for a model with  $B_{p}=5.4\cdot
10^{14}$G are shown in the left-hand panel of Fig.~\ref{fig7}. As expected, the $U_{n}$ and $C_{n}$ QPOs appear at higher 
frequencies than model A2 (see Table~\ref{tab3}),  while between the $U_{n}^{(-)}$ frequencies 
 we notice some smaller amplitude peaks which appear near the crustal modes frequencies of a non-magnetised star. 
From the effective 2D-FFT,  we find that these peaks have an intermediate character between a purely crustal and core mode, and 
therefore we denote them $^{l}t_{n}^{\ast}$. It is important to stress however that they are global oscillations and not crustal modes.    
 In fact,  these $^{l}t_{n}^{\ast}$ oscillations have an eigenfunction pattern with a complex structure.
Perpendicular to the magnetic field lines, the number of nodes  
is identical to an unmagnetised torsional crustal mode $^{l}t_{n}$, while along the field lines,  
it is equal to the next higher frequency $U_{n}$ QPO. 
The origin of their oscillation pattern may be due to the smaller difference between the Alfv\'en and shear velocities at the crust/core boundary in model A3, 
which facilitates  an  energy transfer of the initial crust excitation 
to several open magnetic field lines.  
In Fig.~\ref{fig8}, we show the effective 2D-FFT for a series of eight selected magneto-elastic waves  from $U_{0}^{(-)}$ to  $^{5}t_{1}^{\ast}$. 
As Fig.~\ref{fig7} shows the various $^{l}t_{n}^{\ast}$  oscillations are gradually damped during the evolution.  
An exception is $^{2} t_{0}^{\ast}$ which is the strongest peak also during the later time interval. 
This might be due to its position between the $U_{0}^{(-)}$ and $U_{1}^{(-)}$ --- presumably a continuum gap. 

This suggests that at higher field strengths, more $^{l} t_{0}^{\ast}$
modes may become long-lived. To check this, we also study  a model with  
field $B_{p} = 2.37\cdot 10^{-3} \sqrt{G} \rho_0 R$ ( $B_{p}=1.35 \cdot 10^{15}$G for the fiducial model). 
As shown in the right-hand panel of Fig.~\ref{fig7}, the $U_{n}$ QPOs now have higher frequency  
and the $^{4} t_{0}^{\ast}$ and $^{6} t_{0}^{\ast}$ oscillations  are between the $U_{0}^{(-)}$ and $U_{1}^{(-)}$ frequencies. 
These $^{l} t_{0}^{\ast}$ do indeed seem to last longer than the previous model 
with weaker magnetic field, and suggest the possibility that some of
the low frequency QPOs actually observed may be these magneto-elastic $^{l} t_{0}^{\ast}$ oscillations. 
 Note however that the $^{4} t_{0}^{\ast}$ and $^{6} t_{0}^{\ast}$ oscillations now have both a single node 
along the magnetic field line instead of, respectively, two and three nodes as for the  $B_{p}=5.4\cdot 10^{14}$G case shown in 
Fig.~\ref{fig8}.


At higher frequencies  where  the overtones of non-magnetised shear modes reside we find 
several $^{l}t_{n}^{\ast}$ oscillations similar to what is observed for the fundamental modes. 
This means that the crustal mode overtones are also able to excite
global oscillations, which in the core 
have high numbers of nodal lines. An example is shown in Fig.~\ref{fig8} for $^2 t_{1}^{\ast}$ and 
$^5 t_{1}^{\ast}$ modes, which have, respectively, dimensionless oscillation frequencies $\nu  = 0.0675 \sqrt{ G \rho_0 } $ and 
$ 0.0678 \sqrt{ G \rho_0 } $. In the crust, we can recognise a radial nodal line of an $n=1$ crustal mode which excites a very high overtone of
a core magnetic mode. 
In order to appreciate this effect, a simulation with higher radial resolution is needed. We used in this case a $48\times120$ grid resolution.
This behaviour, already reported in~\cite{2013arXiv1304.3566G},
confirms the different behaviour of QPOs in superfluid neutron stars. 
However, it would be important to determine the duration of these oscillations in order to explain the higher frequency QPOs observed in the tail of SGR 1806-20 with 
625 and 1837~Hz, which, respectively, lasted about 200 and 15~s. 
With the current resolution, we are not able to provide a secure answer to this question.

\subsection{QPO interpretation} \label{sec:QPOid}

The observed QPOs can be roughly divided into a set of low- and high-frequency QPOs. In fact, the majority of the frequencies reside in the
band $\nu < 155 $~Hz, with only two outside this range, at $625$ and
$1837$Hz --- both detected from SGR1806-20. We wish to examine our
results in the context of these observations, to see which model stellar
parameters are required to provide oscillations in the appropriate range. Our goal here is not to determine a model that strictly 
 fits all the observed frequencies, as our stars are still rudimentary
 in many respects. In particular, it would be more appropriate to 
use realistic EoSs instead of polytropes to explore the neutron star
parameter space. This problem will also require a proper description of 
gravity in general relativity and will be addressed in future work. 
With our current models, we can however understand the effects of composition stratification and entrainment on the spectrum and their influence 
on the magnetic field strength required to explain the observations. 

For our models we consider  a standard mass $M=1.4 M_{\odot}$ and vary the radius in order to tune the QPOs' frequency range. This choice will 
determine at the same time the transformation relation from the dimensionless to the physical units for the frequency and magnetic field strength. 

\begin{table*}
\begin{center}
  \caption{\label{tab4} 
Potentially observable magneto-elastic oscillations from model S.  The frequencies are given in Hz  for a model S with $M=1.4M_{\odot}$  and
 $R=10$~km. The averaged magnetic field and its value at the pole are given in gauss in the second and third columns, respectively.  These frequencies  may be rescaled for a star with a different radius by using 
 $\nu \sim R^{-3/2}$.} 
\begin{tabular}{  c  c  c c c c c c c c c c c  }
  \hline 
 Model   &  $\langle B \rangle$ & $B_p$    & $C_0$ & $C_1$ & $C_2$ & $a^{(+)}$ & $a^{(-)}$ & $U_{\ast}$ & $U_0^{(-)}$ &  $U_0^{(+)}$ & $U_1^{(-)}$ & $U_1^{(+)}$   \\
  \hline
  S     &  $ 1.0 \cdot10^{15}$  & $3.3\cdot10^{14}$  &$15.4$ & 29.9 & 44.7 &  5.7 &  11.4 &  7.0 & 12.8 & 18.7 & 25.6  &  32.6 \\
    S   & $2.5\cdot10^{15}$    & $8.3 \cdot 10^{14}$  & 38.4 & 74.3 & 111.4 & 11.3 & 24.6 & 14.6 & 29.9 & 42.7 & 59.6 & 76.9 \\ 
\hline
\end{tabular}
\end{center}
\end{table*}

Let us start with model S and consider a radius $R=10$~km and two possible magnetic field strengths, namely $B_{p} = 3.3\cdot10^{14}$G and $B_{p} = 8.3\cdot10^{14}$G, 
which, respectively,  correspond to an average magnetic field
$\langle B \rangle  = 10^{15}$G and $\langle B \rangle  =
2.5\cdot10^{15}$G.  Note that in model S, the field is more buried
than for model A, with an interior field about three times that at the
polar cap. Already for these intermediate magnetic fields,  
the magneto-elastic waves of model S penetrate the crust in both the open and closed field line regions. This is a crucial property in the current paradigm for the interpretation 
of QPOs as a result of the modulation of a trapped fireball by crustal vibrations \citep{1995MNRAS.275..255T}. 
 
As Table~\ref{tab4} shows, even at a relatively low field of $B_{p} =
3.3\cdot10^{14}$G many $C_{n}$ and $U_{n}$ QPOs provide an approximate
match to the observations: for instance 
$U_0^{(+)} \simeq 18$Hz,  $U_1^{(-)}  \simeq 26$Hz,  $C_1  \simeq
30$Hz, while the higher overtone of $U_{n}$ frequencies can describe the
other QPOs at $<155$Hz.  A similar identification  can be  found with 
model S and $B_{p} = 8.3\cdot10^{14}$G. 
This numerology is, of course, a little
arbitrary at this stage --- one might instead choose to associate our
lowest-frequency mode, $a^{(+)}$, with the lowest-frequency reported
QPO: the $16.9$Hz frequency found from a recent re-analysis of the
SGR1806-20 data \citep{2011A&A...528A..45H}.

For model A3 the QPO frequency range is better described by a star
with $R=14$~km.  In contrast with model S, the $C_{n}$ oscillations are confined in the core for  intermediate 
magnetic fields, but due to the strong entrainment at the bottom of the inner crust many $^l t^{*}_{0}$ oscillations appears at low frequencies and live longer. 
 In Table~\ref{tab5} we show the QPO frequencies for two magnetic
 fields, $ B_p = 2.8 \cdot10^{14}$G and  $B_p = 6.9 \cdot 10^{14}$G.  For the latter case,  
 the $U_{n}$ oscillations penetrate the crust and hence are suitable
 for association with observed QPOs. In this case however we see from Table~\ref{tab5} that the $U_{n}$ oscillations 
 cannot explain the very lowest frequency QPO ($16.9$~Hz). An
 intriguing possibility to explore for model A3 is the case of $B_p  >
 10^{15}$G and a smaller shear modulus.   The spectrum at low frequencies might be populated more by $^l
t^{*}_{0}$ oscillations, while the $U_{n}$ QPOs would be shifted at higher
frequencies. Many of the low QPOs could be interpreted with a series
of $^l t_{0}^{\ast}$ oscillations similar to those already proposed
initially for unmagnetised models.  In particular, it would be
interesting to investigate this  possibility by adding the effect of magnetic
fields to the analysis of~\cite{2009PhRvL.103r1101S}.

From these preliminary results it is evident that in superfluid magnetar models the poloidal magnetic field required to describe the QPO spectrum is weaker than in normal fluid stars. 
This is  an expected outcome which comes from the different definition of the  Alfv\'en speed in these two systems. 
Another interesting feature of superfluid models could be the identification of the two high-frequency QPOs. In the standard single-fluid model, these two high frequency QPOs have 
been hard to explain, due to the resonant absorption of the highly
dense magnetic core
oscillations~\citep{2011MNRAS.410.1036V,2012MNRAS.420.3035V}.  In the
superfluid case, the higher frequency QPOs could represent overtones of the $^l t_{n}^{\ast}$ 
oscillations, as recently proposed by~\cite{2013arXiv1304.3566G}: the smaller difference between the Alfv\'en and shear speed 
at the crust/core interface should facilitate a resonance between the crust and the core oscillations. For instance for model A3 with $ \langle B \rangle = 10^{15}$G we find that 
the $^l t_{1}^{\ast}$ may vary  within $(490, 811)$Hz,  $^l t_{2}^{\ast}$ within $(877, 1453)$Hz and $^l t_{3}^{\ast}$ in  $(1243 , 2059)$Hz. The lower and upper limits of the oscillation frequencies 
correspond to stellar models with, respectively, $R=14$ and 10~km.  
In order to assess the duration of these oscillations and therefore their relevance for the QPO identification one requires a high-resolution numerical grid, as at these frequencies the 
core oscillations have a large number of nodal lines. The solution of
this question is beyond our current numerical capabilities, but we hope to
return to it in the future. 
\begin{table*}
\begin{center}
  \caption{\label{tab5} 
Potentially observable magneto-elastic oscillations from models A2 and
A3.  The frequencies are given in Hz  for models with $M=1.4M_{\odot}$  and
 $R=14$~km. As in the last table, the averaged magnetic field and its value at the pole are given in gauss  in the second and third column, respectively. These frequencies  may be rescaled for a star with a different radius by using 
 $\nu \sim R^{-3/2}$.   The upper index  $ ^{\star}$ denotes oscillations which are damped  in $t< 1 s$, while the $^{\dagger}$ indicates magneto-elastic waves which do not penetrate the crust.
  } 
\begin{tabular}{  c  c c c  c  c   c c c c c c  }
  \hline 
 Model   &  $\langle B \rangle$ & $B_p$  & ${}^2t_0^\ast$ & ${}^3t_0^\ast$ & ${}^4t_0^\ast$  & $C_0$  & $U_{\ast}$ & $U_0^{(-)}$ &  $U_0^{(+)}$ & $U_1^{(-)}$ & $U_1^{(+)}$   \\
  \hline
  A2   & $5.3 \cdot10^{14}$  &  $2.8 \cdot10^{14}$ &     $49.0^{\star}$ & $-$    & $104.6^{\star}$ & $13.5^{\dagger}$   &   9.1$^{\dagger}$   & 18.8$^{\dagger}$ & 28.5$^{\dagger}$  & 38.7$^{\dagger}$  & 47.4$^{\dagger}$  \\
  A3  &  $5.3 \cdot10^{14}$  &  $2.8 \cdot10^{14}$     & 31.9   & 47.8   &    $-$    & 15.1     &  10.3$^{\dagger}$ & 24.4$^{\dagger}$ &  36.7$^{\dagger}$ & 49.2$^{\dagger}$   & 61.9$^{\dagger}$ \\
  A3  &  $1.3 \cdot10^{15}$   & $6.9 \cdot 10^{14}$   & 33.6   & 40.5 & 72.5    & $-$  & 26.7 & 64.0 & 94.2 & 123.9  &  148.6 \\
\hline
\end{tabular}
\end{center}
\end{table*}

Another possibility is that the high-frequency QPOs are
non-axisymmetric magnetic modes of the superfluid core. For surface
fields of the order of $10^{15}$ G, we found that the 150 and 625 Hz
QPOs of SGR1806-20 and the 155 Hz QPO of SGR1900+14 could plausibly be
`fundamental' $m=2$ Alfv\'en oscillations
~\citep{2013MNRAS.429..767P}. Even in this scenario, however, the 1837 Hz frequency would have
to be some kind of higher order mode --- either an overtone or a mode
with $m>2$.

\section{Discussion} \label{sec:discussion}

The original discovery of magnetar QPOs, in the right frequency
range to be shear modes of the crust, brought hopes of probing
the neutron star interior and constraining the EoS of
dense matter. These hopes have been somewhat diluted, as it has become
clear that more physics is at play in magnetar QPOs than simply the star's elastic
crust. On the other hand, these complications mean that there is potentially additional
information to be gleaned from these observations: the magnetic field
structure within a neutron star and the nature of any
superfluid/superconducting components.

The entirety of observations to date consist of two sets of data: the
decaying X-ray tails that followed the giant flares of two soft
gamma repeaters. These tails are clearly modulated at the rotation
rate of each star; it has been suggested that what we see is a
fireball above the star but anchored to its surface \citep{1995MNRAS.275..255T}. If the observed
QPOs do indeed correspond to oscillations of the star, these must also
be propagated from the surface through this fireball. The QPOs
themselves cover a frequency band from tens of Hz up to kHz, many of
them only appearing in certain parts of the X-ray tail.

In this work, we attack the problem of modelling magnetar QPOs by looking at the behaviour of
linearised perturbations of a magnetised star, accounting for the
effect of superfluid neutrons and an elastic crust. This clearly only
addresses part of the phenomenology described above --- indeed, the
best-studied part of it --- attempting to link the frequencies of
observed QPOs to different classes of stellar
oscillation. Nonetheless, we can still learn something from this
endeavour.

In a magnetar with a single-fluid crust and two-fluid core  with composition stratification,
we find that many magnetic oscillations are able to penetrate the crust and
could explain the lowest frequency observed QPOs, 17-18Hz, as
well as others in the range up to $\sim 150$ Hz. 
In a magnetar model with superfluid neutrons both in the inner crust and the core, the strong crustal 
entrainment reduces the frequencies of the shear modes.  As a result, a new 
series of  magneto-elastic waves with hybrid character appears. 
  For a typical magnetar field strength ($ B_{\p} \sim 10^{15}$G)
they have frequency close to that of 
shear modes of a unmagnetised model, but  their oscillation pattern is not localised into the crust.   We call these 
global oscillations $^l t_{n}^{\ast}$ in order to recall their analogy with crustal axial modes (torsional modes).

In superfluid models with strong entrainment in the crust and $B_{\p} \sim 10^{15}$G,  the frequency range $25\lesssim\nu\lesssim 150$~Hz contains a
number of  $^l t_{n}^{\ast}$ modes which are not damped by coupling
with the core, in contrast with the shear waves of the single-fluid case. Moreover,  the overtones of 
$^l t_{n}^{\ast}$ oscillations  appear at higher frequencies and  could explain QPOs in the range
$150\lesssim\nu\lesssim 2000$ Hz.  This result is in agreement with the recent work of~\citet{2013arXiv1304.3566G} 
Numerical limitations however prevent us from reliably determining their duration.  
In our current models with superfluid neutrons in the crust, QPOs below $\sim 25$ Hz
become harder to explain  if a `standard' shear modulus value  and 
 typical magnetar field strengths are used. 
 
 Recent results from~\citet{2013arXiv1304.3566G} find discrete
  Alfv\'en modes in a superfluid star, replacing the continuous
  spectrum of single-fluid models. In an attempt to confirm this
  result, we have studied also the behaviour of the QPO phase in our
  stellar models to determine the discrete/continuum nature of the
  spectrum.  Unfortunately, our results do not provide conclusive
  evidence that the spectrum is fundamentally different in the
  superfluid case. This is certainly a problem which  deserves more 
   accurate investigation in future work.

Our current Newtonian polytropic models 
are able to describe the properties of many magnetar QPOs, even with
a relatively low dipolar magnetic field, depending on the particular
model. It would be possible to `tune' our results to the observed data and
find a fit for each observed frequency.  Nonetheless, with so many
degrees of freedom at play, we do not think any reliable
conclusions could be drawn from this approach. Instead, a more
detailed analysis of the parameter space will be carried out in a
future work where also relativistic effects and  
tabulated EoSs will be considered.

Important physics typically missing from studies of the
dynamical-timescale problem of linearised oscillations include the
effect of proton superconductivity and the fact that magnetars may not
be old enough for neutrons to have condensed into a superfluid
throughout the core, but only in regions. Like the effect of 
entrainment, this would tend to reduce the superfluid enhancement of
Alfv\'en frequencies.  For example, one might naively assume that
magnetic oscillations of SGR1900+14 should be of lower frequency than
the corresponding oscillations of SGR1806-20, since the former
magnetar has a lower inferred dipole field. It is also an order of
magnitude older \citep{2012ApJ...761...76T}, however, so its core should contain a larger
region with superfluid neutrons --- and its oscillations would then be
\emph{higher} in frequency.

Perhaps a more pressing problem is explaining why some QPOs
appear and disappear in the middle of the tail but others survive for
hundreds of seconds. This might involve studying possible mode
couplings to see if some oscillations maintain their energy at the
expense of others. More detailed modelling is clearly also required to
explain the locally heated, post-flare magnetosphere of the magnetar
and how stellar oscillations propagate through this to become visible
to us.

\section*{Acknowledgements}

AP acknowledges support from the European Union Seventh Framework Programme (FP7/2007-2013) under grant agreement no. 267251 ``Astronomy Fellowships in 
Italy (AstroFIt)''.  SL  acknowledges support from the German Science Foundation (DFG) via SFB/TR7. 
We would like to thank A. Colaiuda, M. Gabler,  K. Kokkotas, L. Stella and N. Stergioulas for fruitful discussions and comments.

\appendix

\section{Wave equation coefficients}

We write here the coefficients of the wave equation~(\ref{eq:wv}):  
\begin{align}
 A_{1} &= \mu + \frac{B_{r}^{2}}{4\pi} \, , \\
 A_{2} & = \frac{\mu}{r^2} + \frac{B_{\theta}^{2}}{4 \pi r^2} \, , \\
 A_{3}  & = \frac{B^{r} B^{\theta }}{2 \pi r} \, , \\
 A_{4} & = \frac{2 \mu}{r} + \frac{d \mu}{d r} +  \frac{1}{4 \pi r}  \left[ B^{\theta} \partial_{\theta} B^{r}  
- B^{r} \left(  2  B^{r} + \partial_{\theta}  B^{\theta}    \right. \right.  \nn \\ 
 &  \left. \left.  +   \cot \theta  B^{\theta} \right) \right] \, ,  \\
 A_{5} & = \frac{\mu \cot \theta }{r^2} +  \frac{1}{4 \pi r^2}  \left[ B^{\theta} \left( \partial_{\theta} B^{\theta} - B^{r} \right)   
+ r B^{r}  \partial_{r}  B^{\theta}   \right]  \, , \\
 A_{6} & = - \frac{\mu}{r^2 \sin^2 \theta} - \frac{1}{r} \frac{d \mu }{dr } + \frac{1}{4 \pi r^2}  \left[ B^{r} \left(  2 B^{r} + \partial_{\theta} B^{\theta}     \right. \right.  \nn \\ 
 &  \left. \left.  -  r \cot \theta \partial_{r} B^{\theta} \right)   
+  B^{\theta} \left( B^{\theta} -  \partial_{\theta}  B^{r} -\cot \theta \partial_{\theta} B^{\theta} \right)   \right]  \, .
\end{align}
In the limit of zero shear modulus $\mu = 0$, the quantities $A_{k}$ become the coefficients of the wave equation for the core's protons.


\nocite*
\bibliographystyle{mn2e}

\label{lastpage}

\end{document}